\documentclass{aa}  
\usepackage{lscape}
\usepackage{graphicx}
\begin{document}

   \title{Isocyanogen formation in the cold interstellar medium}


   \author{C. Vastel
          \inst{1}
          \and
          J.C. Loison\inst{2}
          \and
          V. Wakelam\inst{3}
          \and
          B. Lefloch\inst{4}
          }

   \institute{IRAP, Universit\'e de Toulouse, CNRS, UPS, CNES, F-31400 Toulouse, France
              \email{cvastel@irap.omp.eu}
              \and
             Institut des Sciences Mol\'eculaires (ISM), CNRS, Univ. Bordeaux, 351 cours de la Lib\'eration, 33400, Talence, France
             \and
             Laboratoire d'astrophysique de Bordeaux, Univ. Bordeaux, CNRS, B18N, all\'ee Geoffroy Saint-Hilaire, Pessac, 33615, France
             \and
             CNRS, IPAG, Univ. Grenoble Alpes, F-38000 Grenoble, France
             }

   \date{Received January 3 2019; accepted March 23 2019}

 
  \abstract
   {Cyanogen (NCCN) is the simplest member of the dicyanopolyynes group, and has been proposed as a major source of the CN radical observed in cometary atmospheres. Although not detected through its rotational spectrum in the cold interstellar medium, this very stable species is supposed to be very abundant. }
   {The chemistry of cyanogen in the cold interstellar medium can be investigated through its metastable isomer, CNCN (isocyanogen). Its formation may provide a clue on the widely abundant  CN radical observed in cometary atmospheres.}
   {We performed an unbiased spectral survey of the L1544 proto-typical prestellar core, using the IRAM-30m and have analysed, for this paper, the nitrogen chemistry that leads to the formation of isocyanogen. We report on the first detection of CNCN, NCCNH$^+$, C$_3$N, CH$_3$CN, C$_2$H$_3$CN, and H$_2$CN in L1544. We built a detailed chemical network for NCCN/CNCN/HC$_2$N$_2^+$ involving all the nitrogen bearing species detected (CN, HCN, HNC, C$_3$N, CNCN, CH$_3$CN, CH$_2$CN, HCCNC, HC$_3$N, HNC$_3$, H$_2$CN, C$_2$H$_3$CN, HCNH$^+$, HC$_3$NH$+$) and the upper limits on C$_4$N, C$_2$N. The main cyanogen production pathways considered in the network are the CN~+~HNC and N~+~C$_3$N reactions.}
   {The comparison between the observations of the nitrogen bearing species and the predictions from the chemical modelling shows a very good agreement, taking into account the new chemical network. The expected cyanogen abundance is greater than the isocyanogen abundance by a factor of 100. Although cyanogen cannot be detected through its rotational spectrum, the chemical modelling predicts that it should be abundant in the gas phase and hence might be traced through the detection of isocyanogen. It is however expected to have a very low abundance on the grain surfaces compared to HCN. }
{}
   \keywords{astrochemistry--molecular processes--line: identification--molecular data--radiative transfer}

   \maketitle
%

\section{Introduction}

Molecular complexity in the cold interstellar medium has been revealed to be extremely rich, due to the sensitivity and spectral resolution of modern instruments. Organic molecules have been detected, such as alcohols, aldehydes, acids, ethers, amines, and nitriles. The strong bond of the cyano group (--C$\equiv$N) is present in many species such as cyanopolyynes (HC$_n$N with n=3,5,7, ...), and it has also been speculated that dicyanopolyynes might be abundant in interstellar clouds \citep{kolos2000,petrie2003}. Dicyanopolyynes are very stable species formed by highly unsaturated linear chain of carbon atoms ended by a cyano group at each edge, such as N$\equiv$C--(C$\equiv$C)$_n$--C$\equiv$N. The simplest form of dicyanopolyynes is the non polar species, cyanogen (NCCN, also called C$_2$N$_2$), that cannot be observed through its rotational spectrum. Its metastable isomer, CNCN has however been detected recently towards L483, which harbours a Class 0 protostar, and tentatively in the cold dark cloud TMC-1 \citep{agundez2018} by stacking of four transitions. The most plausible formation route in cold clouds is the reaction\\
\begin{equation}
\rm CN + HNC \rightarrow NCCN + H
\end{equation}
\begin{equation}
\rm \hspace{1.7cm} \rightarrow CNCN + H
.\end{equation}
The detection of protonated cyanogen (NCCNH$^+$) in TMC-1 by \citet{agundez2015} is also a proof that cyanogen might be an abundant species in the interstellar medium, produced through the protonation of NCCN with H$_3^+$ and HCO$^+$:\\
\begin{equation}
\rm XH^+ + NCCN \rightarrow NCCNH^+ +X
\end{equation}
with X=H$_2$ and CO. 
NCCN, which has been detected in Titan atmosphere \citep{teanby2006} but not  within cometary comae, has been proposed as a major source of the CN radical observed in cometary atmospheres \citep{bockelee1985,bonev2000}. Indeed, the photolysis of HCN, HC$_3$N, or NCCN could explain the CN origin. However the HC$_3$N production rate is likely too low to reproduce CN density profile and the CN distribution in comets cannot be explained by the HCN photolysis only \citep{fray2005}. The presence of NCCN in comets is therefore a reliable hypothesis to explain the CN origin. CNCN is a good probe to trace the presence of NCCN, since the two are supposed to be linked through reactions (1) and (2). \\

Some of the precursors of the molecules detected within comets are very likely formed within molecular clouds \citep[see, for example,][]{geiss1999,bockelee2000,ehrenfreund2000}, and understanding of the formation of CNCN and NCCN under cold cloud conditions is a crucial step       s the understanding of the molecular complexity in comets. As part of the IRAM-30m Large Program ASAI\footnote{Astrochemical Surveys At Iram: http://www.oan.es/asai/} \citep{lefloch2018}, we carried out a highly sensitive, unbiased spectral survey of the molecular emission of the L1544 prestellar core with a high spectral resolution. In the present study we have examined the nitrogen chemistry in this core using the observations of isocyanogen, protonated cyanogen, and cyanoethynyl, used a radiative transfer modelling to determine the observed column densities and compared with the most up-to-date chemical modelling for cyanogen chemistry. \\

In Section 2 we present the observations from the ASAI spectral survey and the line identification for the cyanogen bearing species. Based on the detections and tentative detection, we compute in Section 3 the column densities of these species. In Section 4 we present a new chemical modelling and confront the results with the observations. 

\section{Observations}

\begin{figure}
   \centering
   \includegraphics[width=0.85\hsize]{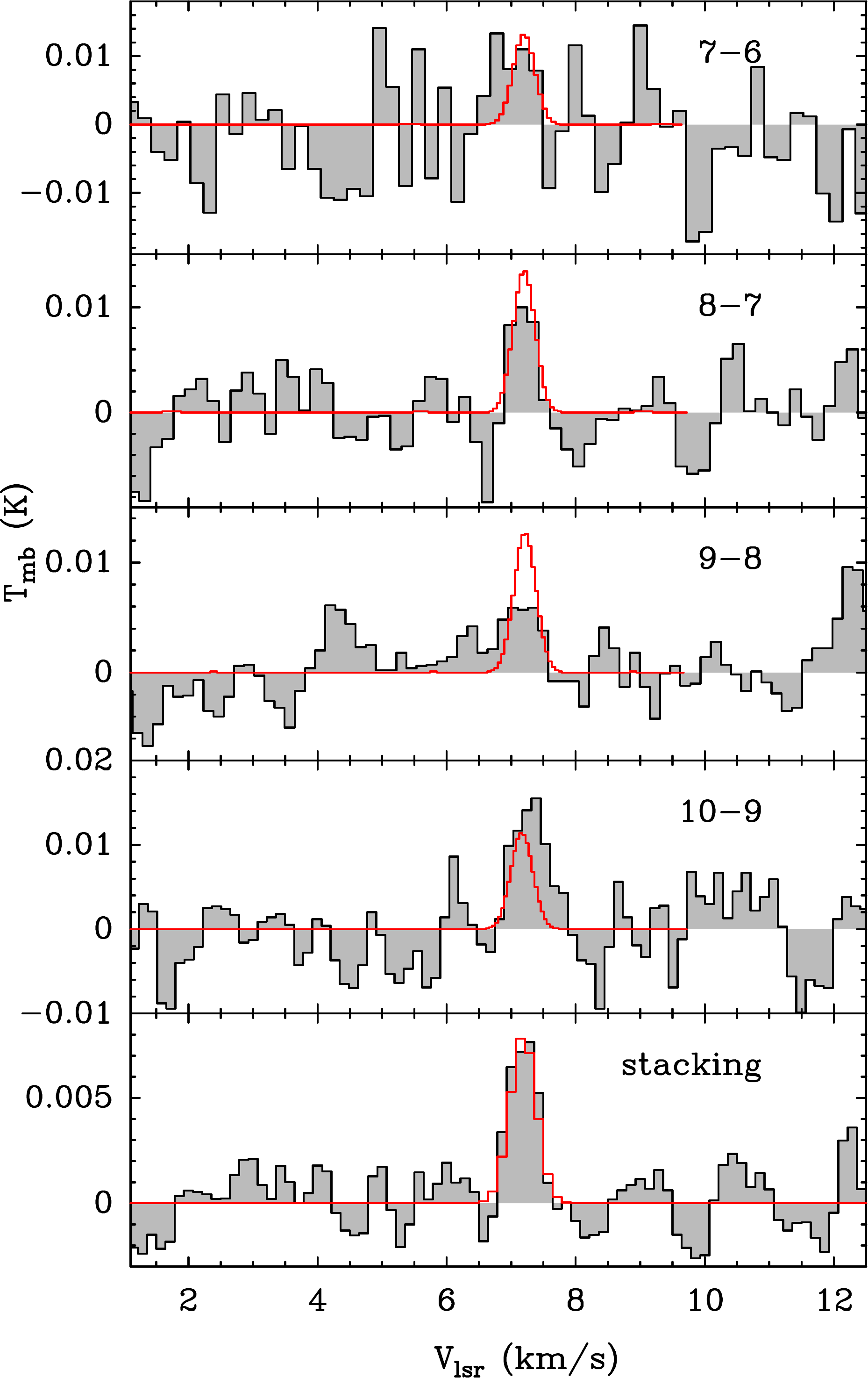}
   \caption{Observed transitions of isocyanogen (CNCN) in black with a LTE model (red) using T$\rm_{ex}$~=~15 K and N~=~6.5~$\times$~10$^{11}$ cm$^{-2}$. Bottom panel shows the results (in black) from the stacking of the four observed transitions as well as the Gaussian fit (in red).}
   \label{CNCN}
 \end{figure}

\begin{figure}
   \centering
   \includegraphics[width=0.85\hsize]{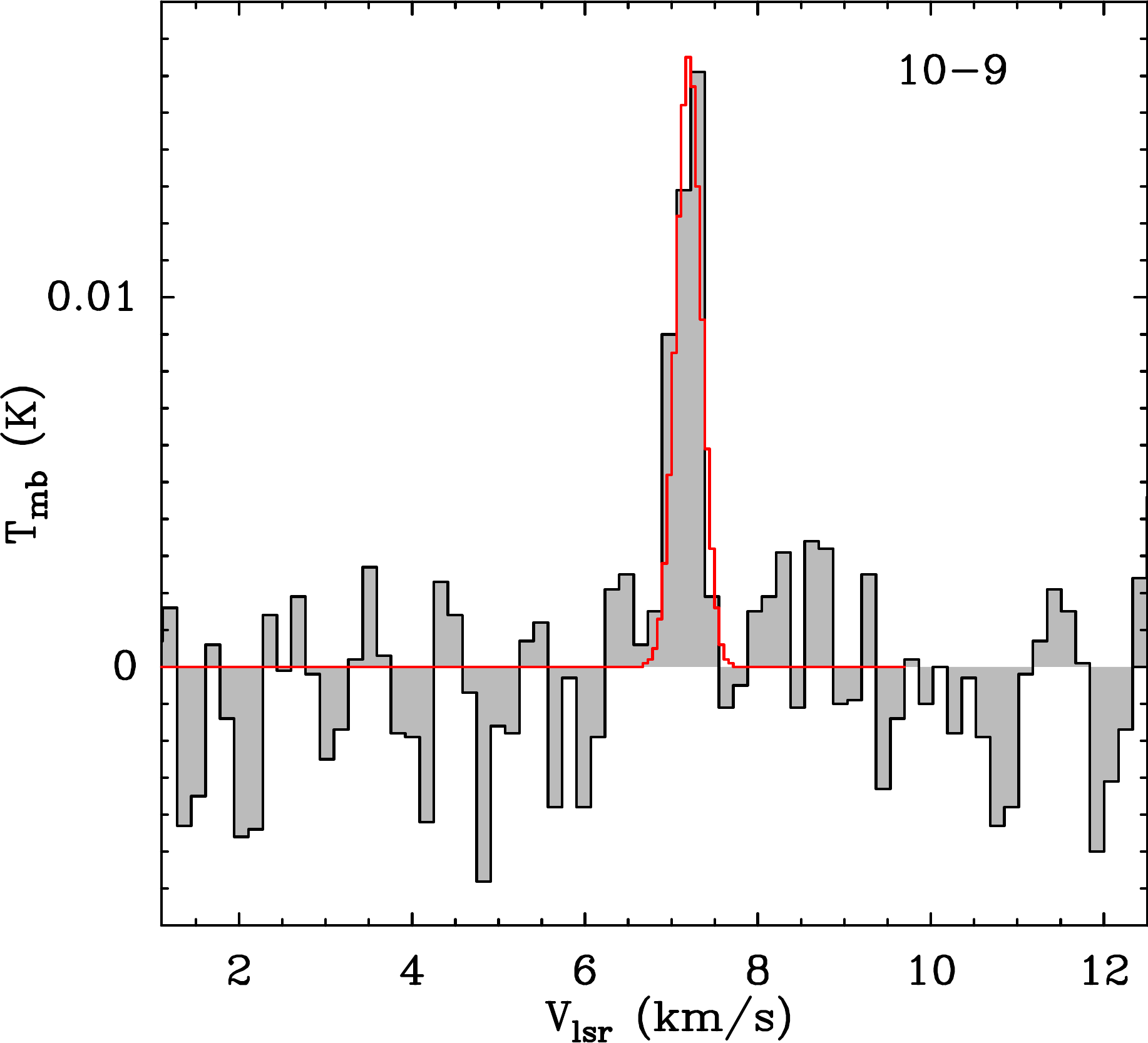}
   \caption{Observed transition of protonated cyanogen (NCCNH$^+$) with a LTE model (red) using T$\rm_{ex}$ = 10 K, and N = 1.9~10$^{10}$ cm$^{-2}$.}
   \label{HCCNHP}
 \end{figure}

\begin{figure}
   \centering
   \includegraphics[width=0.85\hsize]{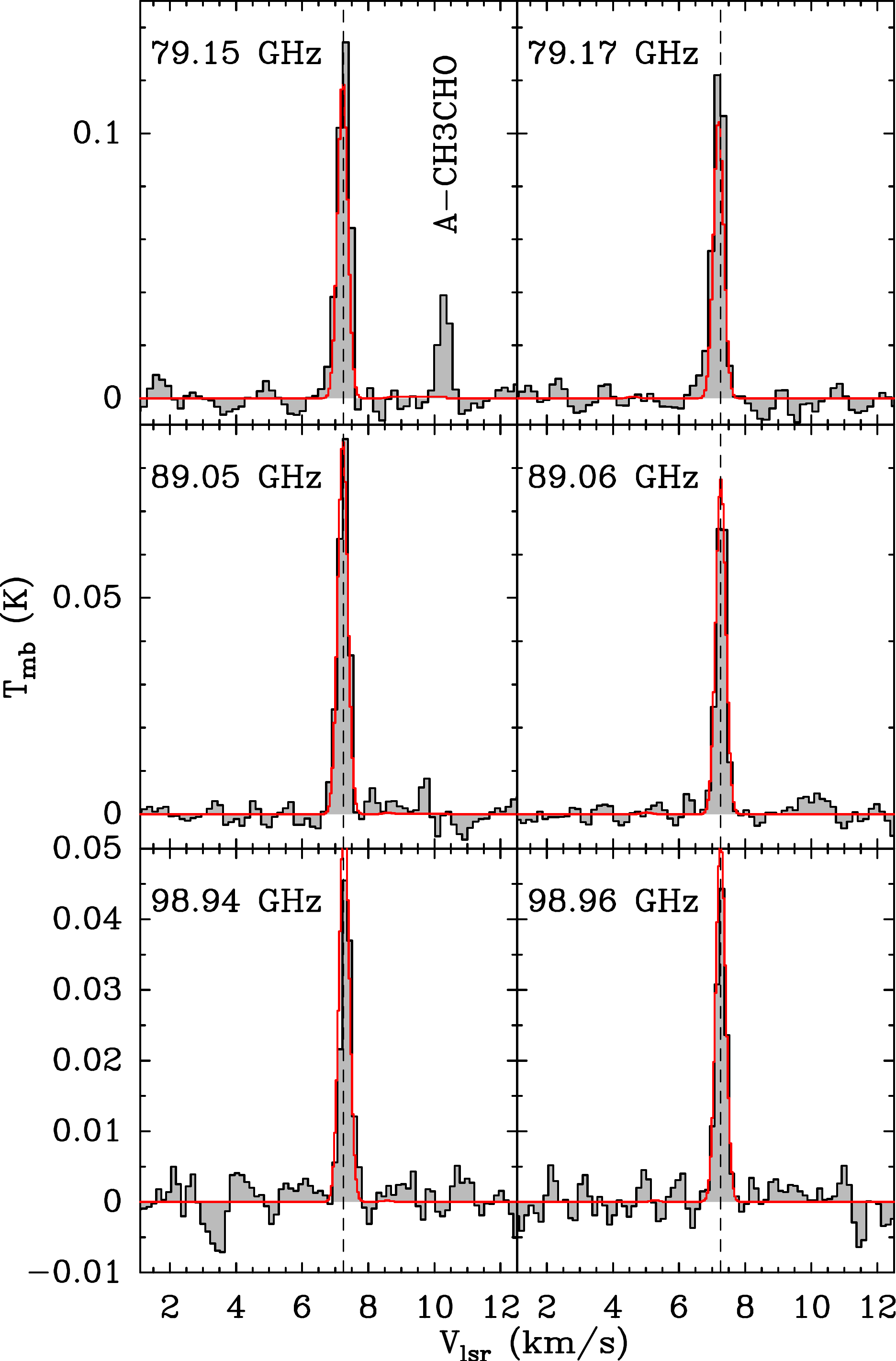}
   \caption{Observed transitions of cyanoethynyl (C$_3$N) with an associated V$\rm_{LSR}$ of 7.25 km~s$^{-1}$ (dashed lines) with a LTE model (red) using T$\rm_{ex}$~=~7.2 K and N~=~1.6~$\times$~10$^{12}$ cm$^{-2}$.}
   \label{C3N}
 \end{figure}

\begin{figure}
   \centering
   \includegraphics[width=0.85\hsize]{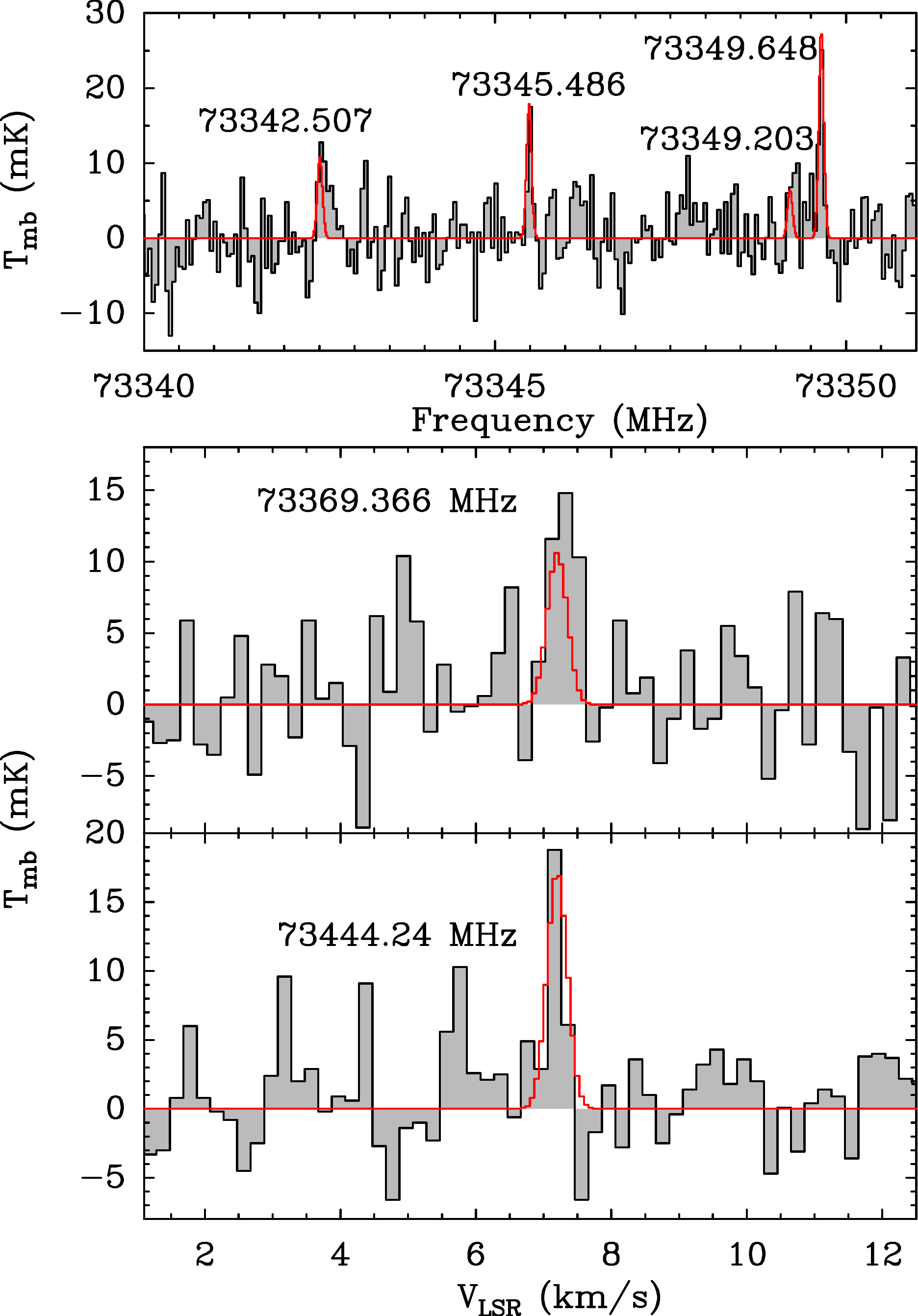}
   \caption{Observed transitions of methylene amidogen (H$_2$CN) with a LTE model (red): T$\rm_{ex}$~=~10 K, N~=~3~10$^{11}$ cm$^{-2}$, $\Delta$V~=~0.35 km~s$^{-1}$.}
   \label{H2CN}
 \end{figure}

\begin{figure}
   \centering
   \includegraphics[width=0.85\hsize]{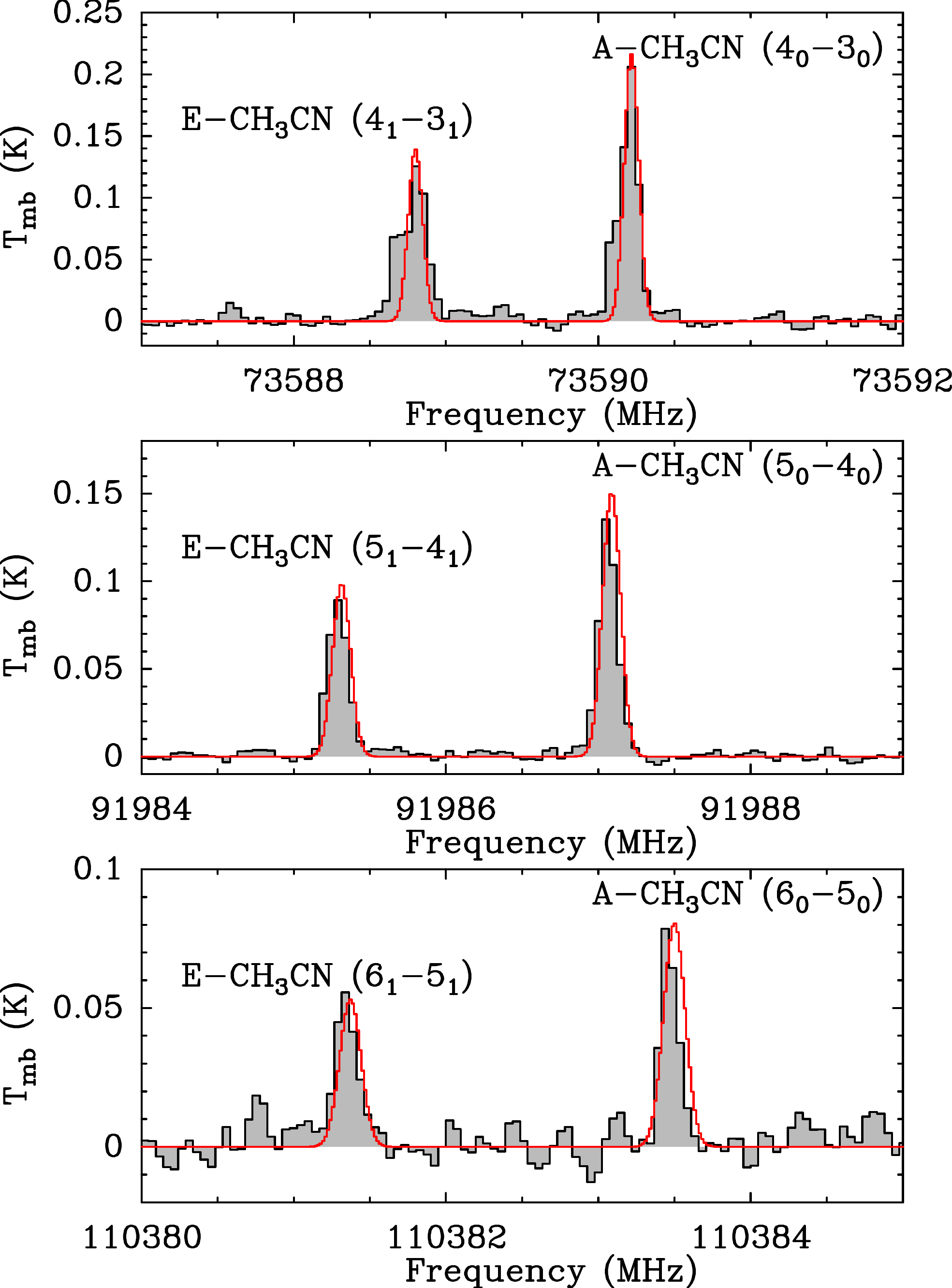}
   \caption{Observed transitions of methyl cyanide (CH$_3$CN) with a LTE model (red): T$\rm_{ex}$~=~5K, N(A--form)~=~4.5~10$^{11}$ cm$^{-2}$, N(E-form)~=~2.8~10$^{11}$ cm$^{-2}$, $\Delta$V~=~0.45 km~s$^{-1}$.}
   \label{CH3CN}
 \end{figure}

\begin{figure}
   \centering
   \includegraphics[width=1.0\hsize]{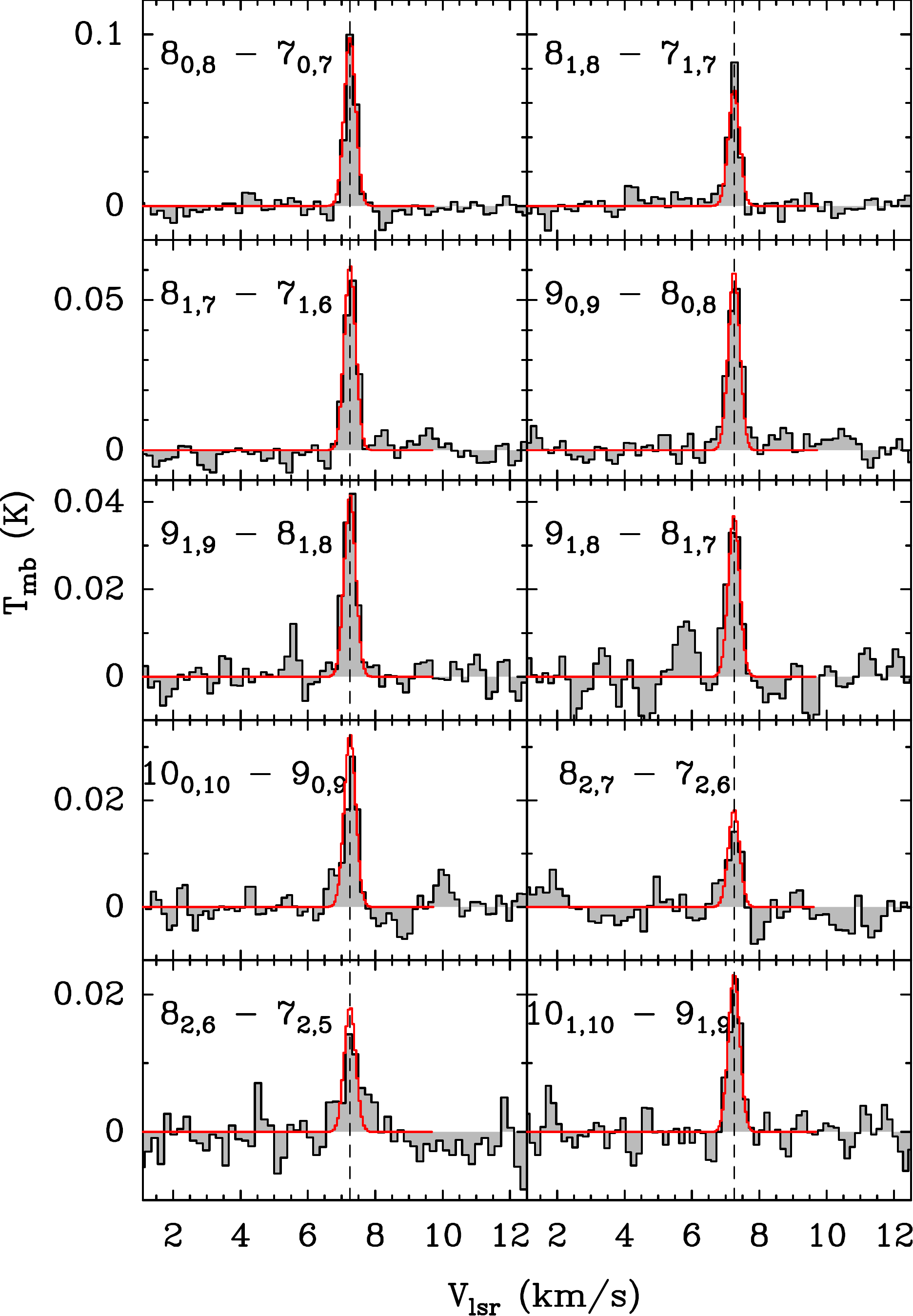}
   \caption{Observed transitions of C$_2$H$_3$CN with an associated V$_{LSR}$ of 7.25 km~s$^{-1}$ (dashed lines) with a LTE model (red): T$\rm_{ex}$~=~5.3 K, N~=~4.0~10$^{12}$ cm$^{-2}$, $\Delta$V=0.4 km~s$^{-1}$.}
   \label{C2H3CN}
 \end{figure}

The observations for all transitions quoted in Table \ref{spectro} were performed at the IRAM-30m towards the dust peak emission of the L1544 pre-stellar core (${\rm \alpha_{2000} = 05^h04^m17.21^s, \delta_{2000} = 25\degr10\arcmin42.8\arcsec}$) in the framework of the ASAI Large Program. Observations at frequencies lower than 80 GHz have been performed in December 2015. All the details of these observations can be found in \citet{vastel2014} and \citet{quenard2017a} and line intensities are expressed in units of main-beam brightness temperature. The frequency range covered by the present spectral survey is 71.74--115.87 GHz. We selected transitions with an upper level energy lower than 30 K, and computed the rms over a range of 15 km~s$^{-1}$ with a spectral resolution of 50 kHz. Some transitions are labelled multiplet or triplet, meaning that the observations do not have the spectral resolution necessary to resolve these transitions or that the hyperfine structure transitions have the same frequencies and slightly different Einstein coefficients. For example, the CNCN 8--7 transition can be decomposed in nine transitions (hyperfine structure due to the quadrupole of the two $^{14}$N nuclei) at the same frequency of 82.784691 GHz:  8$_{7,8}$--7$_{6,7}$, 8$_{9,10}$--7$_{8,9}$, 8$_{8,8}$--7$_{7,7}$, 8$_{9,9}$--7$_{8,8}$, 8$_{9,8}$--7$_{8,7}$, 8$_{8,7}$--7$_{7,6}$, 8$_{7,7}$--7$_{6,6}$, 8$_{8,9}$--7$_{7,8}$, 8$_{7,6}$--7$_{6,5}$. Another example is the 8 (J=17/2)--7 (J=15/2) transition of C$_3$N that can be decomposed in three hyperfine transitions: F=19/2--17/2 at 79.1510059 GHz, F=17/2--15/2 at 79.1509881 GHz and F=15/2--13/2 at 79.1509863 GHz.\\
The line identification and analysis have been performed using the {\sc cassis}\footnote{http://cassis.irap.omp.eu} software \citep{vastel2015a}. The results from the line fitting take into account the statistical uncertainties accounting for the rms (estimated over a range of 15 km~s$^{-1}$ for a spectral resolution of 50 kHz).\\
We report the first detections of CNCN (Isocyanogen), NCCNH$^+$ (protonated Cyanogen), C$_3$N (Cyanoethynyl), H$_2$CN (N-Methaniminyl or Methyleneamidogen), A- and E-CH$_3$CN (Methyl cyanide also known as Acetonitrile or Cyanomethane), and C$_2$H$_3$CN (Vinyl cyanide, also known as acrylonitrile or propenenitrile, H$_2$C=CH--C$\equiv$N) in the L1544 prestellar core shown in Fig. \ref{CNCN} to \ref{C2H3CN} respectively. Some transitions of C$_2$N (Cyanomethylidyne) and C$_4$N (Cyanopropynylidyne) are covered, although not detected, in this spectral survey. \\
Only the CNCN J= 8-7 and 10-9 transitions can be considered as detected (4.2 $\sigma$ for the latter). We have superimposed in red our LTE best fitting solution (see Sect. 3) upon the observed transitions in Fig. \ref{CNCN}. The agreement between the predicted intensities and the observations is satisfying except for the J=9--8 transition. In order to improve the signal to noise ratio of our CNCN detection and the physical parameter determination, we have stacked the four transitions found in the spectral survey. The result is shown in the bottom panel of Fig. \ref{CNCN} (in black, sampled to the resolution of the 10--9 transition) and shows an emission line (in red) detected at a 7$\sigma$ confidence over 12 km~s$^{-1}$ (7.7$\sigma$ over 5 km~s$^{-1}$), with the following parameters for the Gaussian fit: T=8 mK, FWHM=0.48 km~s$^{-1}$, V$\rm_{LSR}$=7.19 km~s$^{-1}$. The CNCN detected transitions (8--7 and 10--9) present different line widths quoted in Table. \ref{spectro} (0.41 km~s$^{-1}$ and 0.61 km~s$^{-1}$). These do, however, have high uncertainty on the V$\rm_{LSR}$ of the 8--7 transition.  
A similar difference, if confirmed, has been observed for the detection of CNCN in the L483 dense cloud (line widths between 0.47 and 0.87 km~s$^{-1}$) \citep{agundez2018}. \\
A feature seems to be present in the spectrum of the 9--8 transition (Fig. \ref{CNCN}) at about 12 km~s$^{-1}$. The unidentified (using the JPL and CDMS databases) transition is also seen in the spectrum of L483 \citep{agundez2018} at $\sim$ 5 km~s$^{-1}$ with respect to the 9--8 transition. The selection of the baseline around 5 km~s$^{-1}$ is therefore necessary to avoid this feature, hence a 7.7$\sigma$ detection for the stacking (see above). \\
We will compute, in the following section, the column densities for the detected species and some upper limits.

\begin{sidewaystable*}
\tiny
\caption{Properties of the observed transitions for E$_{up}$ lower than 30 K. The spectroscopic parameters are from JPL for CNCN \citep{gerry1990} and CDMS \citep{muller2005} for NCCNH$^+$ \citep{amano1991}, C$_3$N \citep{gottlieb1983}, CH$_3$CN \citep{cazzoli2006}, C$_2$H$_3$CN \citep{muller2008} and H$_2$CN \citep{yamamoto1992}. The rms has been computed over a range of 15 km~s$^{-1}$ with a spectral resolution of 50 kHz. \label{spectro} }
\begin{tabular}{|c|c|c|c|c|c|c|c|c|c|}
  \hline
    Species  & QN & Frequency  & $\rm E_{up} $ & A$_{ij}$   & rms & $\rm T_{mb}$ & W & FWHM & $\rm V_{LSR}$ \\
                  &        & (GHz)  & (K)  & (s$^{-1}$) &  (mK) & (mK) & (mK~km~s$^{-1}$) & (km~s$^{-1}$) & (km~s$^{-1}$)\\
  \hline
CNCN                    & 7 -- 6 & $72.436917$ (multiplet) & $13.91$ & $\ge$ 10$^{-6}$   & $7.7$ & -- &  --   & -- & --  \\
                              & 8 -- 7 & $82.784691$ (multiplet)& $17.88$ & $\ge$ 10$^{-6}$   & $3.6$ & $11.5 \pm 4.9$ &  $5.0 \pm 3.8$   & $0.41 \pm 0.20$ & $7.2 \pm 0.1$  \\
                              & 9 -- 8 & $93.1323$ (multiplet)& $22.35$ & $\ge$ 10$^{-6}$   & $3.1$ & -- &  --   & -- & --  \\
                              & 10 -- 9 & $103.4798$ (multiplet)& $27.31$ & $\ge$ 10$^{-6}$   & $5.3$ & $15.6 \pm 1.5$ &  $10.1 \pm 2.1$   & $0.61 \pm 0.07$ & $7.3 \pm 0.1$  \\
NCCNH$^+$         &  9 -- 8    & $79.8826559$  &  $19.17$  &  $1.17~10^{-4}$  & $5.3$ & -- & --  &  -- & --\\
                             &  10 -- 9    & $88.7581034$  &  $23.43$  &  $1.61~10^{-4}$  & $2.4$ & $16.6 \pm 2.2$ & $7.4 \pm 2.0$  &  $0.42 \pm 0.06$ & $7.2 \pm 0.1$ \\
                             &  11 -- 10    & $97.6334236$  &  $28.11$  &  $2.15~10^{-4}$  & $5.4$ & -- & --  &  -- & -- \\
C$_3$N               &  8 (J=17/2) -- 7 (J=15/2)   & $79.151$  (triplet) &  $17.09$  &  $\sim 2~10^{-5}$  & $3.9$ & $137.8 \pm 4.1$ & $64.3 \pm 4.8$  &  $0.44 \pm 0.02$  & $7.3 \pm 0.1$ \\    
                             &  8 (J=15/2) -- 7 (J=13/2)   & $79.1697$ (triplet) &  $17.10$  &  $\sim 2~10^{-5}$  & $4.4$ & $132.3 \pm 4.3$ & $57.5 \pm 4.7$  &  $0.41 \pm 0.02$  & $7.2 \pm 0.1$ \\       
                             &  9 (J=19/2) -- 8 (J=17/2)   & $89.0456$ (triplet) &  $21.37$  &  $\sim 3~10^{-5}$  & $2.7$ & $87.6 \pm 2.7$ & $35.3 \pm 2.0$  &  $0.38 \pm 0.01$  & $7.3 \pm 0.1$ \\                                
                            &  9 (J=17/2) -- 8 (J=15/2)   & $89.0643$ (triplet) &  $21.37$  &  $\sim 3~10^{-5}$  & $2.2$ & $76.8 \pm 2.2$ & $28.5 \pm 2.4$  &  $0.35 \pm 0.02$  & $7.3 \pm 0.1$ \\
                            &  10 (J=21/2) -- 9 (J=19/2)   & $98.9400$ (triplet) &  $26.11$  &  $\sim 4~10^{-5}$  & $2.8$ & $46.4 \pm 2.4$ & $18.2 \pm 1.9$  &  $0.37 \pm 0.02$  & $7.3 \pm 0.1$ \\
                            &  10 (J=19/2) -- 9 (J=17/2)  & $98.9587$ (triplet) &  $26.12$  &  $\sim 4~10^{-5}$  & $2.5$ & $44.1 \pm 2.2$ & $16.8 \pm 1.8$  &  $0.36 \pm 0.02$  & $7.3 \pm 0.1$ \\
A-CH$_3$CN      &   4$_0$ -- 3$_0$               &  $73.5902185$   & $8.83$  &  $3.17~10^{-5}$  & $4.7$ & $185.9 \pm 6.0$ & $122.2 \pm 7.9$  &  $0.62 \pm 0.02$ & $7.3 \pm 0.1$ \\ 
                           &   5$_0$ -- 4$_0$               &  $91.9870877$   & $13.24$  &  $6.33~10^{-5}$  & $2.1$ & $133.0 \pm 1.9$ & $66.3 \pm 2.35$  &  $0.47 \pm 0.01$ & $7.3 \pm 0.1$ \\ 
                           &   6$_0$ -- 5$_0$               &  $110.3835001$   & $18.54$  &  $1.11~10^{-4}$  & $5.9$ & $77.2 \pm 5.1$ & $29.5 \pm 4.4$  &  $0.36 \pm 0.03$ & $7.3 \pm 0.1$ \\                     
E-CH$_3$CN      &   4$_1$ -- 3$_1$               &  $73.5887995$   & $7.95$  &  $2.97~10^{-5}$  & $4.7$ & $125.3 \pm 5.0$ & $77.0 \pm 9.7$  &  $0.58 \pm 0.05$ & $7.2 \pm 0.1$ \\ 
                            &   5$_1$ -- 4$_1$               &  $91.9853142$   & $12.36$  &  $6.07~10^{-5}$  & $2.1$ & $89.3 \pm 2.1$ & $49.2 \pm 2.1$  &  $0.52 \pm 0.01$ & $7.3 \pm 0.1$ \\ 
                           &   6$_1$ -- 5$_1$               &  $110.3813721$   & $17.66$  &  $1.08~10^{-4}$  & $5.9$ & $53.4 \pm 3.4$ & $27.7 \pm 4.0$  &  $0.49 \pm 0.04$ & $7.3 \pm 0.1$ \\                     
C$_2$H$_3$CN & 8$_{0,8}$ -- 7$_{0,7}$       &  $75.5856915$   & $16.35$  &  $3.44~10^{-5}$  & $4.1$ & $102.8 \pm 4.1$ & $41.6 \pm 3.8$  &  $0.38 \pm 0.02$ & $7.3 \pm 0.1$ \\                     
                           & 8$_{1,8}$ -- 7$_{1,7}$       &  $73.9815530$     & $18.15$  &  $3.18~10^{-5}$  & $4.4$ & $83.2 \pm 4.6$ & $31.0 \pm 3.5$  &  $0.35 \pm 0.02$ & $7.3 \pm 0.1$ \\                     
                           & 8$_{1,7}$ -- 7$_{1,6}$       &  $77.6338240$     & $18.94$  &  $3.67~10^{-5}$  & $2.9$ & $59.0 \pm 2.0$ & $28.3 \pm 2.8$  &  $0.45 \pm 0.03$ & $7.3 \pm 0.1$ \\                     
                           & 9$_{0,9}$ -- 8$_{0,8}$       &  $84.9460000$       & $20.43$  &  $4.92~10^{-5}$  & $2.6$ & $53.3 \pm 2.6$ & $28.4 \pm 3.1$  &  $0.50 \pm 0.03$ & $7.3 \pm 0.1$ \\                     
                           & 9$_{1,9}$ -- 8$_{1,8}$       &  $83.2075051$   & $22.14$  &  $4.56~10^{-5}$  & $2.8$ & $43.1 \pm 2.8$ & $21.1 \pm 2.7$  &  $0.46 \pm 0.03$ & $7.3 \pm 0.1$ \\                     
                           & 9$_{1,8}$ -- 8$_{1,7}$       &  $87.3128120$    & $23.13$  &  $5.27~10^{-5}$  & $5.3$ & $35.8 \pm 4.4$ & $19.4 \pm 5.0$  &  $0.51 \pm 0.07$ & $7.2 \pm 0.1$ \\                     
                           & 10$_{0,10}$ -- 9$_{0,9}$   &  $94.2766360$    & $24.95$  &  $6.76~10^{-5}$  & $2.9$ & $27.3 \pm 2.5$ & $12.8 \pm 2.6$  &  $0.44 \pm 0.05$ & $7.3 \pm 0.1$ \\                     
                           & 8$_{2,7}$ -- 7$_{2,6}$       &  $75.8388620$    & $25.04$  &  $3.26~10^{-5}$  & $2.6$ & $14.2 \pm 2.3$ & $7.7 \pm 2.8$  &  $0.51 \pm 0.10$ & $7.2 \pm 0.1$ \\                     
                           & 8$_{2,6}$ -- 7$_{2,5}$       &  $76.1288810$    & $25.08$  &  $3.30~10^{-5}$  & $2.8$ & $14.8 \pm 2.2$ & $7.1 \pm 3.4$  &  $0.45 \pm 0.15$ & $7.3 \pm 0.1$ \\                     
                           & 10$_{1,10}$ -- 9$_{1,9}$   &  $92.4262500$    & $26.58$  &  $6.30~10^{-5}$  & $2.2$ & $21.6 \pm 1.9$ & $10.1 \pm 1.8$  &  $0.44 \pm 0.04$ & $7.3 \pm 0.1$ \\                     
\hline
                            & N$_{K_{-1}K_{+1}}$ = 1$_{0,1}$--0$_{0,0}$& &&&&&&&\\
                           & J~F$_1$~F        &  &&&&&&&\\
H$_2$CN            & 1/2~3/2~5/2 -- 1/2~3/2~5/2  & $73.4442400$           &  $3.54$     & $8.20~10^{-6}$     & $3.6$ & $19.1 \pm 3.8$  & $5.5 \pm 2.3$  & $0.27 \pm 0.06$  & $7.2 \pm 0.1$ \\
                           &  3/2~1/2~3/2 --  1/2~1/2~3/2   & $73.3693660$         & $3.52$       & $7.56~10^{-6}$     & $4.2$ & $14.0 \pm 3.9$  & $7.3 \pm 4.7$  & $0.49 \pm 0.18$  & $7.3 \pm 0.1$ \\
                           & 3/2~5/2~3/2 --  1/2~3/2~1/2   & $73.3425070$         & $3.54$       & $7.68~10^{-6}$     & $4.1$ & $14.4 \pm 3.9$  & $7.3 \pm 4.8$  & $0.48 \pm 0.19$  & $7.1 \pm 0.1$ \\
                           &  3/2~5/2~5/2 --  1/2~3/2~3/2   & $73.3454860$         & $3.54$       & $8.50~10^{-6}$     & $4.0$ & $17.4 \pm 4.1$  & $5.5 \pm 2.6$  & $0.30 \pm 0.07$  & $7.1 \pm 0.1$ \\
                           &  3/2~1/2~1/2 -- 1/2~1/2~3/2   & $73.3492030$         & $3.52$       & $9.15~10^{-6}$     & $4.1$ & $7.9 \pm 3.9$  & $2.6 \pm 2.2$  & $0.51 \pm 0.20$  & $7.1 \pm 0.2$ \\
                           &  3/1 5/2 7/2--  1/2 3/2 5/2  & $73.3496480$         & $3.54$       & $9.88~10^{-6}$     & $4.1$ & $26.6 \pm 3.1$  & $8.5 \pm 2.1$  & $0.30 \pm 0.04$  & $7.2 \pm 0.1$ \\                    
\hline
\end{tabular} 
\end{sidewaystable*}

\section{Determination of the column densities}

In this section we determine the column densities for the detected species (CNCN, NCCNH$^+$, C$_3$N, A-CH$_3$CN, E-CH$_3$CN, H$_2$CN, C$_2$H$_3$CN)  and upper limits for C$_2$N and C$_4$N, taking into account the uncertainties based on the line fitting and also the absolute calibration accuracy, around 10$\%$ or better depending on the band considered. For that we used the {\sc cassis} software, that interrogates the JPL and CDMS databases, and performs LTE and non LTE (when the collisions files are available) radiative transfer modelling. The number of detected transitions is not sufficient to explore a wide range of excitation temperatures. \\
For NCCNH$^+$, we vary the excitation temperature between 10 K and 15 K, and compute a column density of (1.2--1.9) $\times$ 10$^{10}$ cm$^{-2}$. A lower value for the excitation temperature is not compatible with the upper limit on the 8--9 transition at 79.9 GHz. We note that the rms for both undetected transitions at 79.9 and 97.6 GHz is more than a factor two higher than for the detected transition at 88.8 GHz, therefore the expected intensities (less than 20 mK) are within the noise level.\\
For H$_2$CN, we vary the excitation temperature between 5 K and 15 K (a reasonable range for species detected in L1544), and compute a column density of (3--6) $\times$ 10$^{11}$ cm$^{-2}$.  For isocyanogen (CNCN), the covered energy range is large enough to constrain the excitation temperature between 13 and 17 K, with a computed column density of [6.0--6.6] $\times$ 10$^{11}$ cm$^{-2}$. \\
For cyanoethynyl (C$_3$N), the resulting excitation temperature is (7.2 $\pm$ 0.2) K and column density is (1.6 $\pm$ 0.1) $\times$ 10$^{12}$ cm$^{-2}$.  For C$_2$N and C$_4$N, we can compute upper limits on the column densities of 2 $\times$ 10$^{12}$ cm$^{-2}$ and 10$^{14}$ cm$^{-2}$ respectively. \\
For methyl cyanide (CH$_3$CN), we used the A and E forms provided in the {\sc cassis} database\footnote{http://cassis.irap.omp.eu/?page=catalogs-vastel} and converged on an excitation temperature of 5 $\pm$ 0.5 K and a column density of (4.5$^{+1.0}_{-1.4}$) $\times$ 10$^{11}$ cm$^{-2}$ for the A form and (2.8$^{+0.6}_{-0.9}$) $\times$ 10$^{11}$ cm$^{-2}$ for the E form, with a A/E ratio of $\sim$ 1.6, slightly higher than the ratio found by \citet{minh1993} towards TMC-1 ($\sim$ 1.5).  For C$_2$H$_3$CN, we detected 10 transitions between 16 and 27 K, and used a rotational diagram analysis: T$\rm_{rot}$=(5.3 $\pm$ 0.3) K and N=(4.0 $\pm$ 0.5) $\times$ 10$^{12}$ cm$^{-2}$. We note that for all species, we carefully checked that the modelled transitions of the undetected lines are consistent with the noise level.\\
We present in Table \ref{comparison} a comparison between the observed column densities in the prototypical dark cloud TMC-1 and the values found in the L1544 prestellar core, at a later evolutionary stage. The latter values will be used in Section 4.2 for the chemical modelling of the core. The column densities are comparable for many species, such as CN, CNCN, H$_2$CN, HCN, HNC, HCNH$^+$, despite the fact that TMC-1 is less evolved than L1544, which is showing signs of gravitational contraction in its centre \citep{caselli2012}. For the other species, the column densities are higher in TMC-1, which can be explained by the fact that depletion is more pronounced in the prestellar core L1544. 

\begin{table}
\tiny
\centering
\caption{Comparison of the observed column densities (in cm$^{-2}$) between the TMC-1 dark cloud and the L1544 prestellar core. \label{comparison}}
\begin{tabular}{ccc}
                                & TMC-1 &      L1544    \\
\hline
CN                          &     (2.9 $\pm$ 0.7) $\times$ 10$^{14}$           &   1.2 $\times$ 10$^{14}$\\
                               &     \citet{crutcher1984}             &   \citep{hily-blant2008}\\
C$_3$N                  &  8.2 $\times$ 10$^{12}$                               & (1.6 $\pm$ 0.1) $\times$ 10$^{12}$\\
                              &         \citet{guelin1998}                              &   This paper\\
CNCN                     &   9 $\times$ 10$^{11}$                               & (6.0--6.6) $\times$ 10$^{11}$   \\
                               &                \citet{agundez2018}                              &  This paper       \\
NCCNH$^+$          &   (8.6 $\pm$ 4.4) $\times$ 10$^{10}$               &   (1.2--1.9) $\times$ 10$^{10}$\\
                              &         \citet{agundez2015}                             &     This paper \\
H$_2$CN               &         1.5 $\times$ 10$^{11}$                  & (3--6) $\times$ 10$^{11}$\\
                              &         \citet{ohishi1994}                               & This paper \\
CH$_3$CN            &   5 $\times$ 10$^{12}$                    & (5--8.9) $\times$ 10$^{11}$\\
                              &                 \citet{minh1993}                                                         & This paper\\
CH$_2$CN            &           (2--10) $\times$ 10$^{13}$              & (0.28--24.6) $\times$ 10$^{12}$\\
                               &                \citet{irvine1988}                      &  \citet{vastel2015b}\\                
HCN                       &      2.5 $\times$ 10$^{14}$                 &  (2.7--4.4) $\times$ 10$^{14}$\\
                              &          \citet{hirota1998}                              & \citet{quenard2017a}\\
HNC                       &             4.2 $\times$ 10$^{14}$                  & (2--4) $\times$ 10$^{14}$\\
                              &         \citet{hirota1998}                              & \citet{quenard2017a}\\
HC$_3$N               &  (1.6 $\pm$ 0.1) $\times$ 10$^{14}$         &   (8.0 $\pm$ 0.4) $\times$ 10$^{13}$   \\
                               & \citet{takano1998}                                     &  \citet{quenard2017a}\\
                               &                                                                   & \citet{hily-blant2018}\\
HNC$_3$               &  (3.8 $\pm$ 0.6) $\times$ 10$^{11}$         &   (0.75--2) $\times$ 10$^{11}$   \\
                               & \citet{kawaguchi1992a}                            &   \citet{vastel2018a}\\
HCCNC                  &  (2.9 $\pm$ 0.2) $\times$ 10$^{12}$        &   (0.85--2.2) $\times$ 10$^{12}$\\
                               & \citet{kawaguchi1992b}                            &   \citet{vastel2018a}\\
HC$_3$NH$^+$     &(1.0 $\pm$ 0.2) $\times$ 10$^{12}$         &   (1--2) $\times$ 10$^{11}$ \\
                               & \citet{kawaguchi1994}                             &   \citet{quenard2017a}\\
HCNH$^+$             &   1.9 $\times$ 10$^{13}$       &  (2 $\pm$ 0.2) $\times$ 10$^{13}$\\
                             &          \citet{schilke1991}           &    \citet{quenard2017a}         \\
C$_2$H$_3$CN    &      3 $\times$ 10$^{13}$      &     (4.0 $\pm$ 0.5) $\times$ 10$^{12}$\\
                              &        \citet{turner1999}                   &     This paper\\
C$_2$N               &         not detected                               & $\le$ 2~10$^{12}$\\
C$_4$N                 &         not detected                               & $\le$ 10$^{14}$\\                                                                               
\end{tabular}
\end{table}

\section{Chemical modelling}

\subsection{Investigating the chemistry of dicyanopolyynes species in the interstellar medium}

To build the chemical network for NCCN/CNCN/HC$_2$N$_2^+$ chemistry we start from the work of \citet{agundez2015,agundez2018b} using a general methodology for the study of the chemistry described in details in \citet{loison2014} and \citet{loison2017}. The rate constants used in our study are presented in Table \ref{reactions} and the details of the calculations are shown in the Appendix. They were added to the chemical network of kida.uva.2014 \citep{wakelam2015a} and additional updates from \citet{wakelam2015a,loison2016,hickson2016,wakelam2017,vidal2017,loison2017} \\
The main NCCN production pathways in our network are the CN + HNC and N + C$_3$N reactions. A minor reaction is the N + HCCN reaction which has been neglected in this study because HCCN is not supposed to be abundant in the interstellar medium. Indeed, HCCN has not been detected in TMC-1, with an upper limit on the fractional abundance (with respect to $\rm H_2$) of $\rm 2 \times 10^{-10}$ \citep{mcgonagle1996}, and it has not been included in our network. The  CN + HNC reaction has been studied theoretically by \citet{petrie2004} leading to a high rate constant at low temperature. The N + C$_3$N reaction was already in the {\sc kida} network but producing only CN + CCN \citep{loison2014}. For this study, we have performed an extensive theoretical calculations on the C$_3$N$_2$ triplet surface showing that N($^4$S) + C$_3$N(X$^2\Sigma^+$) is correlated not only to CN(X$^2\Sigma^+$) + CCN(X$^2\Sigma$) but also to C($^3$P) + NCCN (X$^1\Sigma^+$) and to C($^3$P) + CNCN(X$^1\Sigma^+$) as shown on Table A.1 and Figures A.1, B.1, and C.1 in the Appendix. Considering the exoenergicities, the N + C$_3$N $\rightarrow$ C + NCCN channel should be favoured.\\

Apart from the minor channel N + C$_3$N $\rightarrow$ C + CNCN, the other CNCN production is through the electronic dissociation rate (DR) of HC$_2$N$_2^+$, the latter being produced through NCCN + H$_3^+$ and NCCN + HCO$^+$ reactions (see Equation 3). Other HC$_2$N$_2^+$ production pathways show a barrier as CN + HCNH$^+$ (see Table G.1 in the Appendix). By comparison with similar DRs \citep{florescumitchell2006,plessis2012,reiter2010},  the H elimination seems to be an important pathway, assumed to be equal to 50 $\%$ in this study. In that case, the NCCN production is highly exothermic and will carry a large internal energy, up to 668 kJ/mol (see Table \ref{reactions}). If the internal energy is above the dissociation limit, it will lead to dissociation producing H + CN + CN with a low branching ratio neglected here. On the contrary, if the internal energy is below the dissociation limit but above the isomerization barrier which is the most probable case as hydrogen atom and NCCN will cary some translational energy, it will lead to the production of NCCN and CNCN isomers as shown on Figure D.1 and Table D.1 of the Appendix. The branching ratio is assumed to be proportional to the ro-vibrational density of states of each isomer at the transition state (TS) energy \citep{herbst2000}. As the DR of HC$_2$N$_2^+$ and CNCNH$^+$ will lead to very similar products (NCCN, CNCN, HCN, HNC, and CN) with very similar branching ratios for our assumptions, we consider only one isomer for HC$_2$N$_2^+$ and CNCNH$^+$.\\

Ionic reactions are efficient NCCN and CNCN destruction pathways, either directly through reaction with He$^+$ \citep{raksit1984} or through protonation as the DR of HC$_2$N$_2^+$ does not lead to 100$\%$ of NCCN and CNCN. \citet{safrany1968b} have found some reactivity of NCCN with nitrogen atoms. However, it has been shown that, in this kind of experiment, the reactivity is due to the presence of metastable N$_2^*$ \citep{michael1980,dutuit2013}. We then performed DFT calculations, using the M06-2X/AVTZ method (this highly nonlocal functional developed by \citet{zhao2008} is well suited for structures and energetics of the transition states). Our calculations show that the N + NCCN and N + CNCN reactions have very large barriers in the entrance valley as shown on Tables E.1 and F.1 of the Appendix and are negligible at low temperatures. \\
We have also performed calculations on the N and O reactions with CNCN  and  the H, N, O reactions with NCCN, see Table E.1 and F.1 of the Appendix showing in all cases large barrier in the entrance valley and then negligible rates at low temperatures.  \citet{safrany1968a} and \citet{whyte1983} deduced from a complex experiment that carbon atoms quickly react with NCCN at room temperature, with a rate constant equal to $\rm 3 \times 10^{-11} cm^3~molecules^{-1}~s^{-1}$. Their experiments are complex but are reliable on the fact that carbon atoms react with NCCN, leading directly or indirectly to CN radicals. However, the products assumed for this reaction, CN + CCN, are widely endothermic as shown of Figure A.1.  The only exothermic exit channel for this reaction, C$_3$ + N$_2$, is spin forbidden. Moreover, C$_3$ and N$_2$ cannot be the only products of the experiments since C$_3$ and N$_2$ have a very low reactivity towards nitrogen atoms contrary to the product obtained in the experiments. They cannot lead to CN which was identified as a product of the C + NCCN reaction in presence of N atoms. \\
An alternative explanation is the formation of NCCCN. In the experiment at few Torr, some, or most, NCCCN adduct will be stabilized through collisions and will react with the nitrogen atoms to produce CN and C$_3$N, as NCCCN has a triplet ground state and is reactive with atoms and radicals. In interstellar molecular clouds the pressure is too low for three-body reactions but radiative association may play a role. \\
Using a model developed in \citep{hebrard2013}, we calculated that $\rm k(T) = 2 \times 10^{-14} \times (T/300)^{-1.5} cm^3~molecules^{-1}~s^{-1}$ for the NCCCN production, leading to a low but non-negligible rate constant at 10 K. There is no experimental or theoretical data on the C + CNCN reaction but considering the very high reactivity of carbon atom with unsaturated hydrocarbons and  HCN and HNC \citep{loison2015}, which have similar -CN and -NC bonds than CNCN, there are very little doubt, if any, that this reaction does not show any barrier. Moreover we do not find any barrier at DFT level (M06-2X/AVTZ) for the C attack on CNCN leading in a first step to CCNCN, CNCCN, and CNCNC adducts. These adducts can evolve in different ways. First there is always the back dissociation into C + CNCN, these back dissociation being important only if other bimolecular exit channels involve high transition states, which is the case of CCNCN. In that case radiative association may also play a minor role neglected here. For CNCCN and CNCNC adducts, the back dissociation will not be the main process as there are pathways leading to C + NCCN channel. As not all the C + CNCN entrance channels lead efficiently (without high TS on the pathway) to bimolecular products, we consider a smaller rate constant that the one given by capture theory for the C+ CNCN $\rightarrow$ C + NCCN reaction. \\

\longtab{
\begin{landscape}
\begin{longtable}{|l|c|c|c|c|c|c|l|}
\caption{Summary of reactions review. Definition of $\alpha$, $\beta$, $\gamma$, F$_0$, g can been found in \citet{wakelam2012} and \citet{wakelam2010}: $\rm k = \alpha \times (T/300)^{\beta} \times exp(-\gamma/T) cm^3~molecule^{-1}~s^{-1}$, T is the temperature and ranges between 10-300 K except for some cases (noted). \label{reactions}}\\
\hline
Reactions  & $\Delta$E & $\alpha$ & $\beta$ & $\gamma$ & F$_0$  & g  & Reference \\
                  &kJ/mol       &                &              &                   &             &      &             \\ 
\hline
\endfirsthead
\caption{continued.}\\
\hline\hline
Reactions  & $\Delta$E & $\alpha$ & $\beta$ & $\gamma$ & F$_0$  & g  & Reference \\
                  &kJ/mol       &                &              &                   &             &      &             \\ 
\hline
\endhead
\hline
\endfoot
$\rm He^+ + C_2N_2 \rightarrow CN^+ + CN + He$    & -437  & $8 \times 10^{-10}$  & 0  & 0  & 3  &  0  & Rate constant from \citet{raksit1984}\\
$\rm \hspace{1.5cm} \rightarrow CCN^+ + N + He$    & -470  & $8 \times 10^{-10}$  & 0  & 0  & 3  &  0  & \\
\hline
$\rm He^+ + CNCN \rightarrow CN^+ + CN + He$    & -522  & $1.6 \times 10^{-10}$  & -0.4  & 0  & 3  &  0  & By comparison with He$^+$ + NCCN taking into account the \\
$\rm \hspace{1.6cm} \rightarrow CNC^+ + N + He$    & -695  & $1.6 \times 10^{-10}$  & -0.4  & 0  & 3  &  0  & dipolar moment of CNCN\\
\hline
$\rm H^+ + C_2N_2 \rightarrow H + C_2N_2^+$    & -22  & $4.0 \times 10^{-9}$  & 0  & 0  & 3  &  0  & Similar to H$^+$ + molecule without dipolar moment, \\
                                                                             &        &                                   &     &     &     &      & IE(NCCN) = 13.37 eV\\
\hline
$\rm H^+ + CNCN \rightarrow H + CNCN^+$    & -65  & 0  &   &   &   &   & Similar to H$^+$ + molecule with dipolar moment. \\
$\rm \hspace{1.5cm} \rightarrow H + C_2N_2^+$    & -99  & $1.0 \times 10^{-8}$  & -0.4  & 0  & 3  &  0  & We avoid introducing CNCN+ which will react quickly \\
$\rm \hspace{1.5cm} \rightarrow NCN^+ + CH$    & +380  & 0  &   &   &  &    & with H$_2$ leading to the same final products that C$_2$N$_2^+$.\\
$\rm \hspace{1.5cm} \rightarrow CH^+ + NCN$    & +154  & 0  &   &   &  &    & \\
\hline
$\rm H + NC_3N \rightarrow HCN + C_2N$    & +15  & 0  &   &   &   &   & NC$_3$N has a triplet ground state so it will react  quickly \\
$\rm \hspace{1.3cm} \rightarrow HNC + C_2N$    & +67  & 0  &   &   &  &    &  with atoms and radicals. The two unpaired electrons are\\
$\rm \hspace{1.3cm} \rightarrow CN + HCCN$    & +108  & 0  &   &   &  &    & localized on central C atom and on terminal N atoms:\\
                                                                          &           &     &    &   &  &    &$N\equiv C-^{\bullet}C=C=N^{\bullet} \leftrightarrow  ^{\bullet}N=C=^{\bullet}C-C\equiv N$  \\
$\rm \hspace{1.3cm} \rightarrow CH + C_2N_2$    & +105  & 0  &   &   &  &    & The $\bullet$ represents the lonely electron and show the reactive sites.\\
\hline
$\rm C^+ + C_2N_2 \rightarrow CNC^+ + CN$    & -65  & $1.0 \times 10^{-9}$  & 0  & 0  & 3  & 0   & Charge exchange is endothermic. Rate constant similar \\
$\rm \hspace{1.35cm} \rightarrow CCN^+ + CN$    & +77  & 0  &   &   &  &    & to C$^+$ + C$_2$H$_2$, close to capture rate.\\
\hline
$\rm C^+ + CNCN \rightarrow CNC^+ + CN$    & -149  & $2.0 \times 10^{-9}$  & -0.4  & 0  & 3  & 0   & Charge exchange is endothermic. Rate constant similar to \\
$\rm \hspace{1.5cm} \rightarrow CCN^+ + CN$    & -8  & 0  &   &   &  &    & C$^+$ + HCN, close to capture rate with temperature\\
                                                                           &        &                                   &     &     &     &      &dependency from charge-dipole interaction. \\
\hline
$\rm C + C_2N_2 \rightarrow NC_3N + h\nu$    & -435  & $2.0 \times 10^{-14}$  & -1.5  & 0  & 10  & 0   & The results from \citet{safrany1968a,safrany1968b} and \citet{whyte1983} \\
$\rm \hspace{1.2cm} \rightarrow C_3 + N_2$    & -294  & 0  &   &   &  &    &are complex, the products assumed for this reaction, CN + CCN, being  \\
$\rm \hspace{1.2cm} \rightarrow CN + CCN$    & +98  & 0  &   &   &  &    &widely endothermic. We have performed theoretical calculations  \\
$\rm \hspace{1.2cm} \rightarrow N + C_3N$    & +157  & 0  &   &   &  &    & (see Appendix). We propose that the experimental observations are \\
$\rm \hspace{1.2cm} \rightarrow C_2 + NCN$    & +266  & 0  &   &   &  &    &due to 3-body NC$_3$N formation. In interstellar molecular clouds \\
                                                                           &             &    &    &   &  &    &the pressure is too low for 3-body reactions but radiative association  \\
                                                                           &             &    &    &   &  &    &may play a role. The radiative association rate constant is \\
                                                                           &             &    &    &   &  &    &calculated using our $\rho$-association model \citep{hebrard2013}.\\
\hline
$\rm C + CNCN  \rightarrow C + C_2N_2$    & -84  & $1.0 \times 10^{-10}$  & 0  & 0  & 3  & 0   & See text and Appendix. \\
\hline
$\rm C + NC_3N  \rightarrow CN + C_3N$    & -142  & $2.0 \times 10^{-10}$  & 0  & 0  & 3  & 0   & NC$_3$N has a triplet ground state so it will react quickly with atoms and   \\
                                                                    &     &        &   &   &  &    &radicals. The two unpaired electrons are localized on central C atom \\
                                                                    &     &        &   &   &  &    &and on terminal N atoms: $N\equiv C-^{\bullet}C=C=N^{\bullet} \leftrightarrow  ^{\bullet}N=C=^{\bullet}C-C\equiv N$\\
\hline
$\rm CH + C_2N_2  \rightarrow C_2N + HCN$    & -81  & $2.0 \times 10^{-10}$  & 0  & 0  & 3  & 0   & There is very likely no barrier for this reaction considering the very high  \\
$\rm \hspace{1.4cm} \rightarrow HCCN + CN$    & +20  & 0  & 0  & 0  &0  & 0   &reactivity of the CH radical.\\
$\rm \hspace{1.4cm} \rightarrow C_3H + N_2$    & -177  & 0  & 0  & 0  &0  & 0   &\\
\hline
$\rm CN + HNC  \rightarrow C_2N_2 + H$    & -103   & $2.0 \times 10^{-10}$  & 0  & 0  & 4  & 0   & See \citet{petrie2004}.  \\
\hline
$\rm CN + HCNH^+  \rightarrow HC_2N_2^+ + H$    & +11   & "0"  &   &   &   &    & The reaction is endothermic and show a barrier of 52 kJ/mol  \\
                                                                                &         &        &   &   &   &    & at M06-2X/AVTZ level  (see Table G.1 of the Appendix)\\
\hline
$\rm N + C_2N_2  \rightarrow C_2N_2 + H$    &    & "0"  &    &    &    &     & Very large barriers for N addition on the nitrogen (91 kJ/mol) or on the   \\
                                                                        &   &       &    &    &    &     & carbon atoms (63 kJ/mol) (M06-2X/AVTZ calculations). The rate   \\
                                                                        &   &       &    &    &    &     & constant value in \citet{safrany1968c} is likely due to the presence   \\
                                                                        &   &       &    &    &    &     & of N$_2^*$ \citep{michael1980,dutuit2013}. \\
\hline
$\rm N + CNCN  \rightarrow C_2N_2 + H$    &    & "0"  &    &    &    &     & The most favourable attack on final carbon atom shows a large barrier for N \\
                                                                        &   &       &    &    &    &     & addition (32 kJ/mol) (M06-2X/AVTZ calculations).\\
\hline
$\rm N + C_3N  \rightarrow C + C_2N_2$        & -157   & $6.0 \times 10^{-11}$  & 0   &  0  &  3  &  0   & Theoretical calculations. Important NCCN formation pathway \\
$\rm \hspace{1.1cm} \rightarrow  C + CNCN$    &  -72   & $1.0 \times 10^{-11}$  & 0   &  0  &  3  &  0   & in dense molecular clouds.\\
$\rm \hspace{1.1cm} \rightarrow CN + C_2N$ &  -43   & $2.0 \times 10^{-11}$  & 0   &  0  &  3  &  0   & \\
$\rm \hspace{1.1cm} \rightarrow C_3 + N_2$  &  -333   & 0  & 0   &  0  &  3  &  0   &  \\
\hline
$\rm N + NC_3N  \rightarrow CN + C_2N_2$        & -299   & $4.0 \times 10^{-11}$  & 0   &  0  &  3  &  0   & NC$_3$N has a triplet ground state so it will react quickly with atoms and radicals. \\
$\rm \hspace{1.3cm} \rightarrow N_2 + C_3N$ &  -337   & $4.0 \times 10^{-11}$  & 0   &  0  &  3  &  0   & The two unpaired electrons are localized on central C atom and on  \\
                                                                        &     &  &  &  &   &    & terminal N atomes $N\equiv C-^{\bullet}C=C=N^{\bullet} \leftrightarrow  ^{\bullet}N=C=^{\bullet}C-C\equiv N$\\ 
\hline
$\rm C_2N_2^+ + H_3^+  \rightarrow HC_2N_2^+ + H_2$ &  -215   & $2.8 \times 10^{-9}$  & 0   &  0  &  1.2  &  0   & See \citet{anicich2003}.\\
\hline
$\rm C_2N_2^+ + HCO^+  \rightarrow HC_2N_2^+ + CO$ &  -67   & $1.7 \times 10^{-9}$  & 0   &  0  &  2  &  0   & By comparison with NCCN + H$_3^+$ considering the \\
                                                                                         &          &                                   &      &      &      &       & H$_3^+$/ HCO$^+$ masses difference.\\
\hline
$\rm CNCN + H_3^+ \rightarrow HCNCN^+  + H_2$ &  -240   & 0  &     &   &     &     & By comparison with NCCN + H$_3^+$ taking into account the low dipole moment  \\
$\rm \hspace{1.5cm} \rightarrow HNCNC^+  + H_2$ &  -256   & 0  &     &   &     &     & of CNCN. We do not introduce HC$_2$N$_2^+$ isomers as their DR will leads to   \\
$\rm \hspace{1.5cm} \rightarrow HC_2N_2+  + H_2$ &     & $2.8 \times 10^{-9}$   & -0.3    & 0  & 2    & 0    & similar products with our statistical calculations.\\
\hline
$\rm CNCN + HCO^+ \rightarrow HCNCN^+ + CO$ & -92 & 0 & & & & &We do not introduce HC$_2$N$_2^+$ isomers as their DR will leads to similar products \\
$\rm \hspace{1.9cm} \rightarrow  HNCNC^+ + CO$ & -109 & 0 & & & & &with our statistical calculations.\\
$\rm \hspace{1.9cm} \rightarrow  HC_2N_2^+ + CO$ &   & $1.6 \times 10^{-9}$ & -0.3 & 0 & 2 & 0 &\\
\hline
$\rm C_2N_2^+ + H_2  \rightarrow HNCCN^+ + H$ &  -200   & 0  &     &    &     &      & \citet{anicich2003}\\
$\rm \hspace{1.35cm} \rightarrow  HC_2N_2^+ + H$ &   & $9.0 \times 10^{-10}$  & 0    &  0  &  1.4   &  0    & \\
\hline
$\rm CNCN^+ + H_2  \rightarrow HNCNC^+ + H$ &  -190   & 0  &     &    &     &      & Same as C$_2$N$_2^+$ + H$_2$. We do not introduce HC$_2$N$_2^+$ isomers \\
$\rm \hspace{1.6cm} \rightarrow  HC_2N_2^+ + H$ &      & $9.0 \times 10^{-10}$  &  0   & 0   &  2   &   0   & as their DR will leads to similar products with our statistical calculations.\\
\hline
$\rm HC_2N_2^+ + e^- \rightarrow H + C_2N_2$  &  -668   & $1.7 \times 10^{-7}$  &     &    &     &      & NCCN/CNCN branching calculated using statistical theory, as well\\
$\rm \hspace{1.5cm} \rightarrow  H +CNCN$ & -583 & $1.0 \times 10^{-8}$  &     &    &     &      & as HNC/HCN ratio.\\
$\rm \hspace{1.5cm} \rightarrow  HCN +CN$ & -616 & $1.0 \times 10^{-7}$  &     &    &     &      & \\
$\rm \hspace{1.5cm} \rightarrow  HNC +CN$ & -564 & $1.0 \times 10^{-7}$  &     &    &     &      & \\
$\rm \hspace{1.5cm} \rightarrow  H + CN +CN$&  -93 & 0  &     &    &     &      & \\
\hline
$\rm HC_3NH^+ + e^-  \rightarrow HC_3N + H$  & -568 & $6.0 \times 10^{-7}$  & -0.58  & 0  & 3 & 0 & The global rate constant and the 52$\%$ of HC$_3$N isomers formation and \\
$\rm \hspace{1.7cm} \rightarrow  HNC_3 + H$ & -369 & $1.4 \times 10^{-7}$  & -0.58    &  0  &  3   &  0    & 48$\%$ of HCN isomers formation is from \\
$\rm \hspace{1.7cm} \rightarrow  H + HCNCC$ & -250 & $1.0 \times 10^{-8}$  & -0.58    &  0  &  3   &  0    & \citet{geppert2004,vigren2012} using also \citet{osamura1999}.  \\
$\rm \hspace{1.7cm} \rightarrow  H + HCCNC$ & -473 & $3.0 \times 10^{-8}$  & -0.58    &  0  &  3   &  0    & The relative HC$_3$N/HNC$_3$/HCNCC/HCCNC and HCN/HNC \\
$\rm \hspace{1.7cm} \rightarrow  C_2H + HCN$ & -377 & $3.6 \times 10^{-7}$  & -0.58    &  0  &  3   &  0    & branching ratios are arbitrary.\\
$\rm \hspace{1.7cm} \rightarrow  C_2H + HNC$ & -345 & $3.6 \times 10^{-7}$  & -0.58    &  0  &  3   &  0    & \\
$\rm \hspace{1.7cm} \rightarrow  C_3N + H + H$ & +29 & 0  & -0.58    &  0  &  3   &  0    & \\
\end{longtable}
\end{landscape}
}

\subsection{Chemistry in the L1544 prestellar core}

We have applied the above chemistry to the prototypical prestellar source L1544. We used the gas-grain chemical code {\sc nautilus} (described in \citet{ruaud2016}) in its three-phase mode (gas-phase, grain-surface, and mantle chemistry) to predict the abundances of the simplest form of dicyanopolyynes species, CNCN, and its related species, CN, NCCNH$^+$, C$_3$N, HCN, HNC, HCNH$^+$, HC$_3$NH$^+$, HC$_3$N, HNC$_3$, HCCNC, H$_2$CN, CH$_3$CN, CH$_2$CN, C$_2$H$_3$CN.\\
The methodology we have adopted here is similar to the studies already performed on L1544 \citep[see for example][]{quenard2017a,vastel2018a,vastel2018b}. A two step model has been used. The first phase represents the evolution of the chemistry in a diffuse or molecular cloud, with T = 20 K and several densities, ranging from 10$^2$ to $2 \times 10^4$ cm$^{-3}$. We present in Table \ref{ci} the values of the initial gas phase elemental abundances, based on the results found by \citet{quenard2017a} and \citet{vastel2018b}. We then used the resulting abundances at 10$^6$ years in a second phase model, taking into account the density and temperature structure of the L1544 prestellar core that shows evidence for gravitational contraction \citep{caselli2012,keto2014}. In order to compare the results from the chemical modelling (abundances with respect to H) with the observations (column density) of the observed species presented in Section 2, we took into account the density profile across the core \citep[see][]{jimenez-serra2016,quenard2017b,vastel2018a,vastel2018b} to determine the column density from the chemical modelling instead of the abundance at a specific radius. This methodology has been used for sources such as prestellar cores where molecular depletion at the centre, due to the low kinetic temperature and high H$_2$ density, affects chemistry. Note that for cold dark clouds such as TMC-1, a single temperature and density have been assumed throughout the cloud, with a H$_2$ column density of $\sim$ 10$^{22}$ cm$^{-2}$ \citep{cernicharo1987,feher2016}. \\
With the modifications of the chemical network presented in Section 4.1 we re-examine the chemistry of nitrogen species such as HCN, HNC \citep{quenard2017a}, HC$_3$N \citep{quenard2017a,hily-blant2018}, HNC$_3$ \citep{vastel2018a}, HCCNC \citep{vastel2018a}, HCNH$^+$ \citep{quenard2017a}, HC$_3$NH$+$ \citep{quenard2017a}, CH$_2$CN \citet{vastel2015b}.\\

\begin{table} 
        \centering
        \caption{Initial gas phase elemental abundances assumed relative to the total nuclear hydrogen density n$_{\textrm{H}}$.\label{ci}}
        \begin{tabular}{lcc}
                \hline\hline
                Species &        \\
                \hline
                He              &       $9.00\times10^{-2}$\\
                N               &       $2.14\times10^{-5}$\\
                O               &       $1.76\times10^{-4}$\\
                C$^+$   &       $7.30\times10^{-5}$\\
                S$^+$   &       $8.00\times10^{-8}$\\
                Si$^+$  &       $8.00\times10^{-9}$\\
                Fe$^+$  &       $3.00\times10^{-9}$\\
                Na$^+$  &       $2.00\times10^{-9}$\\
                Mg$^+$  &       $7.00\times10^{-9}$\\
                Cl$^+$  &       $1.00\times10^{-9}$\\
                P$^+$   &       $2.00\times10^{-10}$\\
                F$^+$   &       $6.68\times10^{-9}$\\
                \hline
        \end{tabular}
\end{table}

\begin{figure*}
        \centering
        \includegraphics[width=1\hsize,clip=true,trim=0 0 0 0]{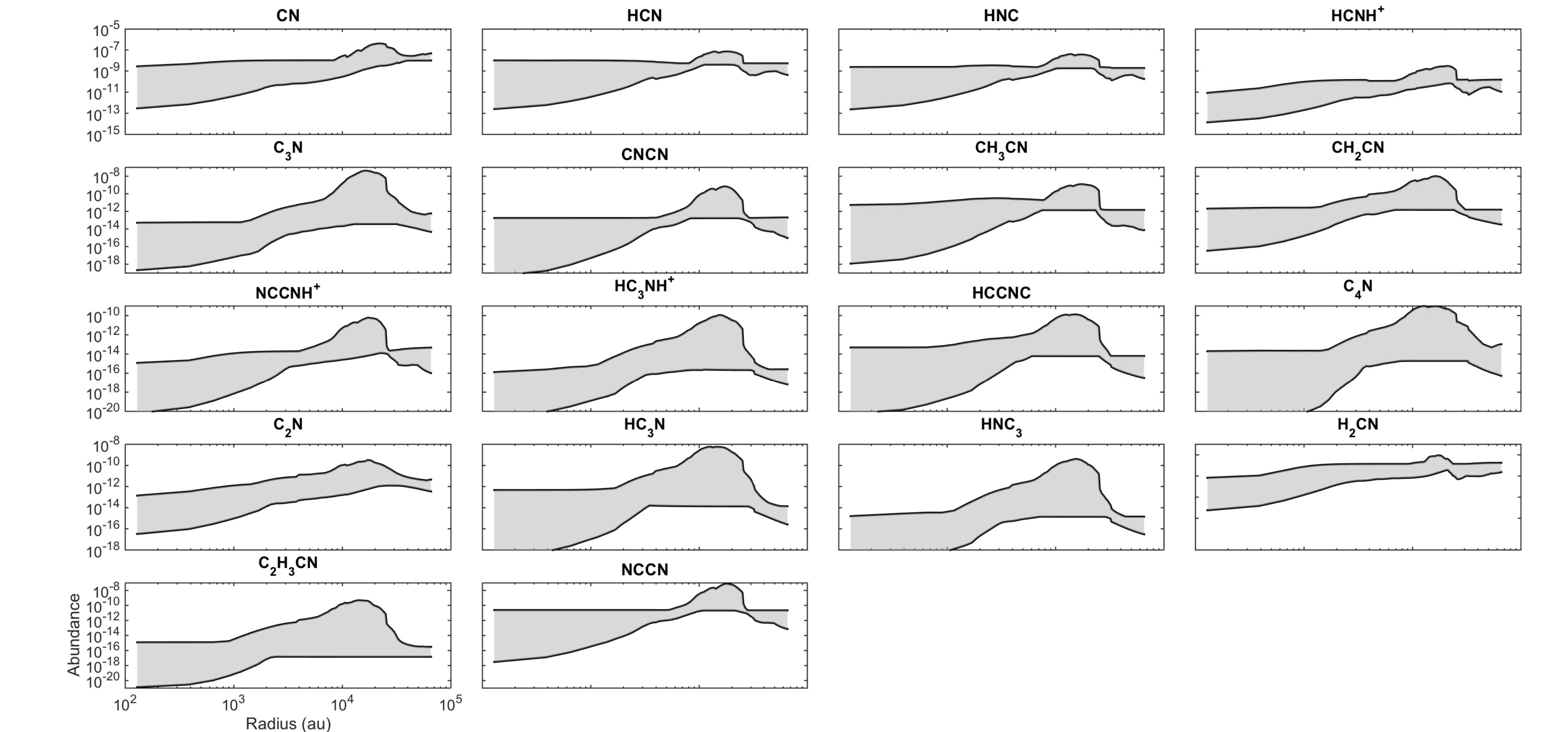}
        \caption{Radial distribution of the nitrogen bearing species abundances in L1544 for an age between 10$^6$ and 10$^7$ years.}
        \label{result_nautilus}
\end{figure*}

\begin{figure*}
   \centering
   \includegraphics[width=1\hsize]{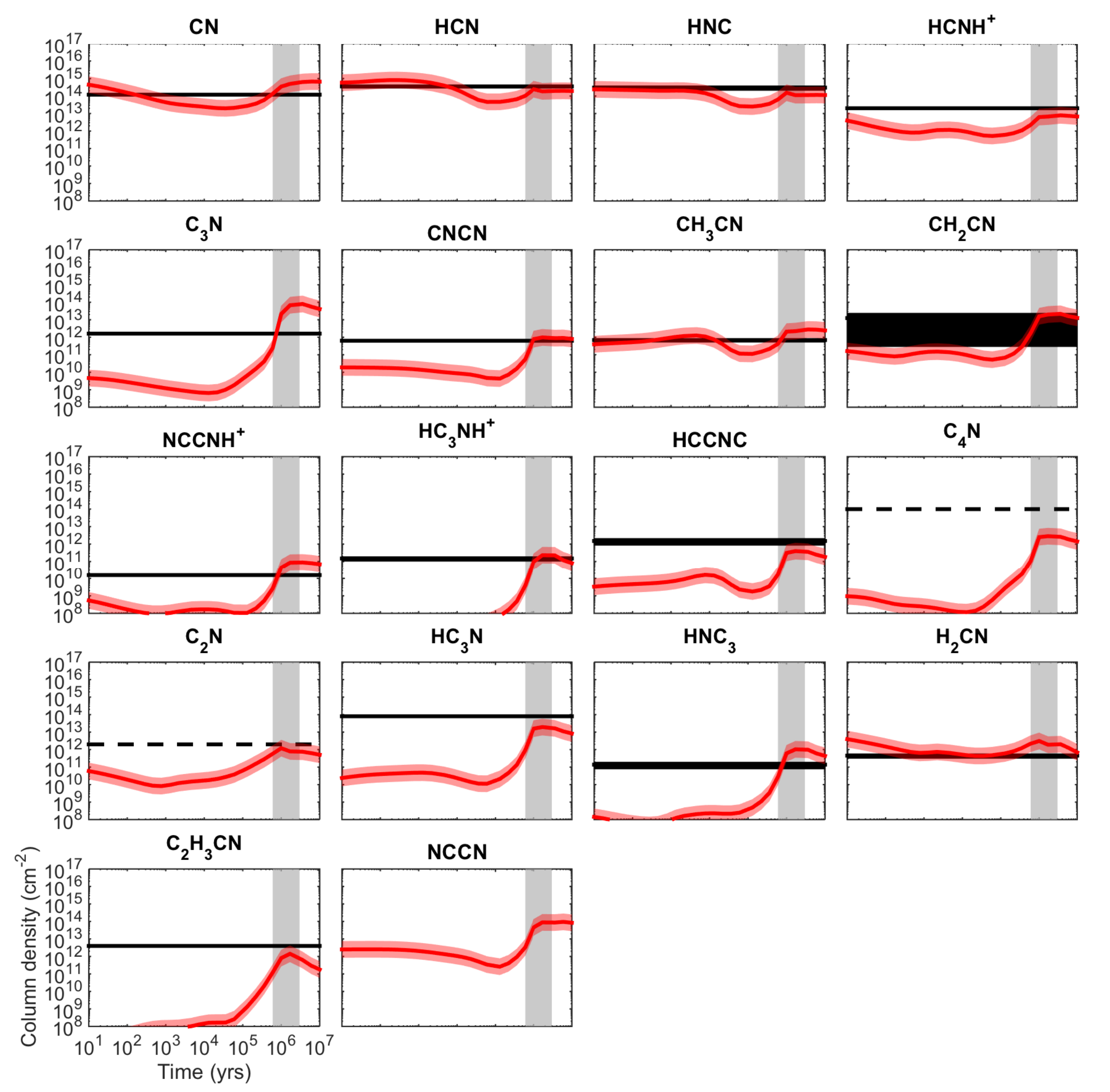}
   \caption{Variation of the modelled column density (in red) as a function of time and the comparison with the observed column density (black horizontal lines). The dashed line corresponds to an upper limit for the column density. The thickness of the black line corresponds to the errors on the column density given in Table \ref{comparison}. A variation by a factor of three on the modelled column density is shown in red. The grey vertical boxes highlight an area between 6 $\times$ 10$^5$ and 3 $\times$ 10$^6$ years (see text).}
   \label{nautilus}
 \end{figure*}

We present in Fig. \ref{result_nautilus} the variation of the modelled abundance as a function of radius for ages between 10$^6$ and 10$^7$ years, for an initial gas density of 3~10$^3$ cm$^{-3}$. We checked that varying the initial density from 10$^2$ to $2 \times 10^4$ cm$^{-3}$ does not change much the results for an age between 10$^6$ and 10$^7$ years. The abundances peak at a radius between [1-2] $\times$ 10$^4$ au, which can be explained by the carbon, oxygen, and sulphur depletion in the cold and dense central regions of prestellar cores. However, this peak is less pronounced compared to other species that do not contain Nitrogen \citep[see for example][for the study of Sulphur bearing species]{vastel2018b}. This can be explained by the fact that nitrogen remains present in the gas phase at densities well above and temperature well below that at which CO depletes on to grains. The resulting molecular differentiation has been seen in many starless cores \citep[e.g.][]{kuiper1996,tafalla1998,ohashi1999,tafalla2006,spezzano2016}.\\ 
We present in Fig. \ref{nautilus} the variation of the modelled column density (in red) as a function of time from {\sc nautilus} and the comparison with the observed column densities (black horizontal lines) and upper limits (dashed horizontal lines) from Section 3. A variation by a factor of three on the modelled column density is shown. The thickness of the black lines corresponds to the uncertainty. 

\begin{figure}
   \centering
   \includegraphics[angle=-90,width=0.8\hsize]{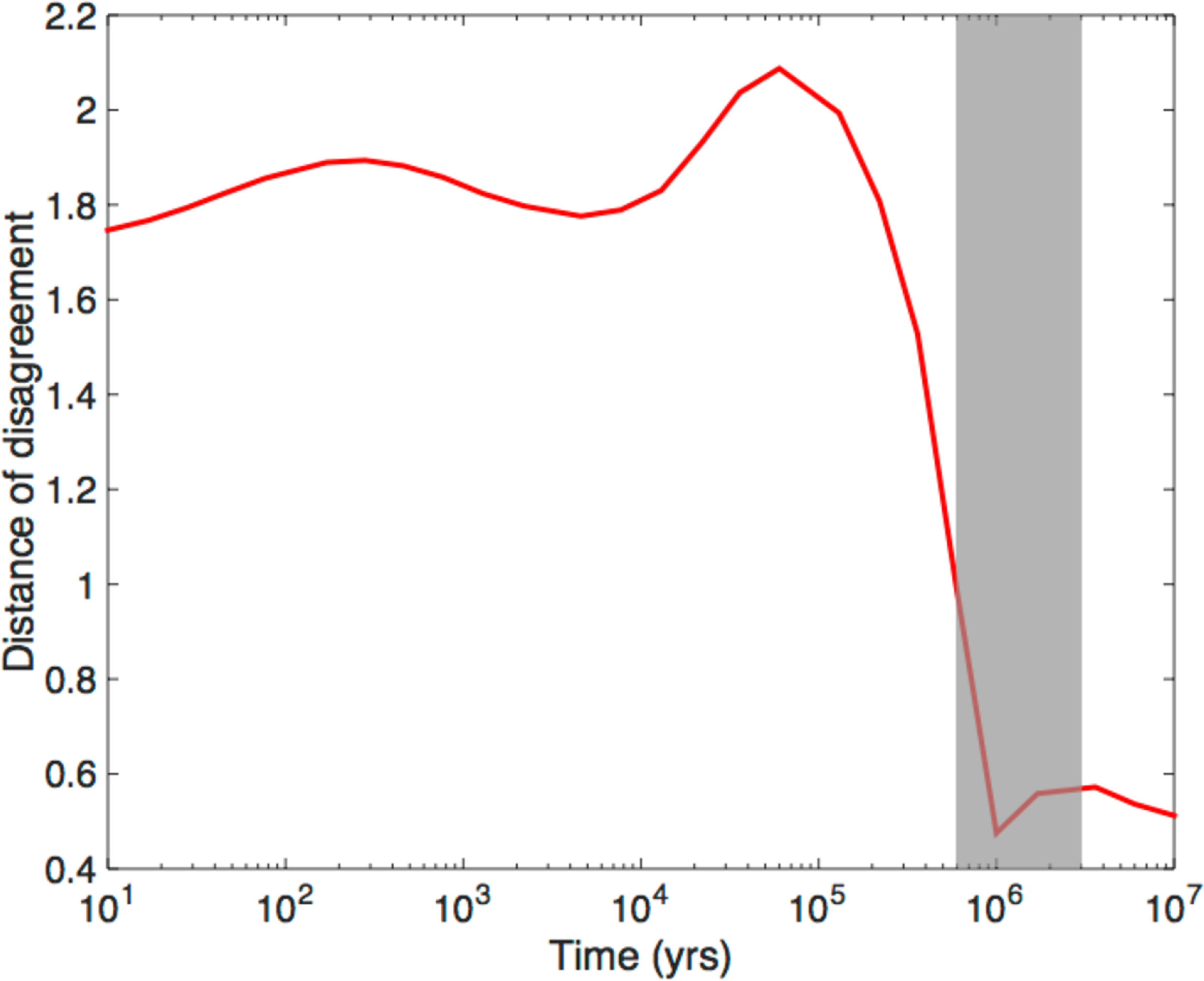}
   \caption{Variation, in time, of the distance of disagreement.The grey box area represent a range between 6 $\times$ 10$^5$ and 3 $\times$ 10$^6$ years (see text).}
   \label{disagreement}
 \end{figure}

To compare the model results with observations, we used the method described in \citet{wakelam2006} in which a distance of disagreement is computed as follows: 
\begin{equation}
\rm D(t) = \frac{1}{n_{obs}} \sum_{i}|log(N(X))_{obs,i}-log(N(X))_{i}(t)|
\end{equation}
where N(X)$_{obs,i}$ is the observed column density, N(X)$_{i}$(t) is the modelled column density at a specific age and n$_{obs}$ is the total number of observed species considered in this computation (14 in the case of nitrogen bearing species detected in L1544). We note that we did not take into account species where a lower limit has been derived from the observations (C$_2$N and C$_4$N).\\
Fig. \ref{disagreement}  shows its temporal variation, where a smaller value corresponds to a better agreement. Taking into account the present study, this distance is lower than unity for times higher than 6~10$^5$ and 3~10$^6$ years. This time is of course model dependent and does not represent the "real" age of the source knowing that the chemical composition is constantly evolving in the interstellar medium. We then used the best-fit age ((1--3) $\times$ 10$^6$ years) from the similar work performed for Sulphur chemistry in L1544 \citep{vastel2018b} and pinpoint a range between $3 \times 10^6$ and $6 \times 10^6$ years for consistency between both studies. \\
We can see from Fig. \ref{nautilus} that all species are reproduced with the present chemical modelling. Using a higher elemental abundance for oxygen and/or carbon (higher by a factor of two, as observed in the diffuse medium, see for instance \citet{jenkins2009}) will over-produce these species, leading to a much higher value for this distance of disagreement. NCCN reaches a calculated gas phase column density around 10$^{14}$ cm$^{-2}$ (and an abundance of $\sim$ 10$^{-8}$ at 10$^4$ au), similar to HCN. From our chemical modelling, the cyanogen/isocyanogen ratio is about a factor of 100. The predicted abundance of NCCN on the grain surfaces remains however low ($\le$ 10$^{-12}$) to be compared to the high grain surface HCN abundance of $\sim$ 5 $\times$ 10$^{-7}$. It can be also noted that in our calculations C$_2$N reaches a non negligible value, close to the upper limit which deserves to thorough investigation with more sensitive observations. The C$_2$H$_3$CN calculated abundance is slightly under estimated which is likely due to the fact that C$_2$H$_4$ is also under estimated in our modelling: C$_2$H$_3$CN is produced through the CN + C$_2$H$_4$ reaction and is a good proxy of C$_2$H$_4$. Then, the gas phase and grain chemistry of C$_2$H$_4$ deserve to be reviewed as C$_2$H$_4$ is an important precursor of complex organic chemistry. \\
Our previous studies \citep{quenard2017a,vastel2018a} tended to overestimate the production of HCN, HNC, HC$_3$N, HNC$_3$, HC$_3$NH$^+$. The inclusion of the new reactions, as well as the various {\sc kida} updates revealed crucial to better understand the nitrogen chemistry, and nitriles in particular, in the cold and dense medium.

The study of the chemical evolution from the prestellar phase to the planetary disk is important to understand how the organics found in our own solar system have been produced. The comparison between the prestellar core phase and the content of the comets studied so far is an important Ariane thread to understand this link. Our study predict a large quantity of NCCN, unfortunately invisible in the rotational transitions tracing the cold medium. This linear symmetric molecule, without permanent dipole moment, has been suggested a long time ago \citep{swings1956} as a likely parent for the production of the CN radical observed in cometary atmospheres. Moreover, NCCN is likely the precursor of cyanomethanimine (HNCHCN), an HCN dimer proposed as precursor of adenine, recently detected by \citet{rivilla2018} towards the Galactic Center quiescent molecular cloud G+0.693. It is likely to have been formed on the surface of icy dust grains through the following reactions \citep{shivani2017}:
\begin{equation}
\rm s-C_2N_2 + s-H  \rightarrow s-NCHCN
\end{equation}
\begin{equation}
\rm  s-NCHCN +s-H \rightarrow s-HNCHCN\\
\end{equation}
It therefore may be at the origin of a more complex cyanogen chemistry. 

\section{Conclusions}

We investigated the formation of isocyanogen in the cold interstellar medium with the study of some nitrogen species in the prototypical prestellar core L1544. We report the detection of isocyanogen (CNCN), protonated cyanogen (NCCNH$^+$), cyanoethynyl (C$_3$N), methylene amidogen (H$_2$CN), methyl cyanide (CH$_3$CN) and acrylonitrile (C$_2$H$_3$CN). We then built a detailed chemical network for the NCCN/CNCN/HC$_2$N$_2^+$ chemistry to understand the isocyanogen formation in the cold interstellar medium. The main cyanogen production pathways considered in this new network are the CN~+~HNC and N~+~C$_3$N reactions. The comparison between the observations of 14 detected nitrogen species in the L1544 prestellar core (CN, HCN, HNC, C$_3$N, CNCN, CH$_3$CN, CH$_2$CN, HCCNC, HC$_3$N, HNC$_3$, H$_2$CN, C$_2$H$_3$CN, HCNH$^+$, HC$_3$NH$^+$) as well as two upper limits (C$_4$N, C$_2$N) and the predictions from the chemical modelling shows a very good agreement. Although NCCN cannot be detected through its rotational spectrum (due to a lack of a permanent electric dipole moment), we can trace this species through the detection of CNCN, since they are both produced through the same reaction. With this new chemical network, we predict a gas-phase column density of $\sim$ 10$^{14}$ cm$^{-2}$ for the "invisible" NCCN, with an expected cyanogen abundance greater than the isocyanogen abundance by a factor of 100. However, its predicted abundance on the grain surfaces remains low compared to the high grain surface HCN abundance. Unless NCCN is efficiently formed within protoplanetary disks, it is an unlikely major source of CN in cometary coma through its photolysis. 

\begin{acknowledgements}
We thank the CNRS program Physique et Chimie du Milieu Interstellaire (PCMI) co-funded by the Centre National d'Etudes Spatiales (CNES). 
\end{acknowledgements}

\begin{appendix}

\section{The C$_3$N$_2$ system}

\begin{table*}[!h]
  \caption{Relative energies at the M06-2X/AVTZ level (in kJ/mol at 0 K including ZPE) with respect to the C + NCCN energy, geometries and frequencies (in cm$^{-1}$, unscaled, calculated at the M06-2X/AVTZ level) of the various stationary points. The absolute energies at the M06-2X/AVTZ level including ZPE in hartree are also given in column 1. The i label represents an imaginary frequency which is characteristic of transition state (TS). The TS is the structure with the highest potential energy along the reaction coordinate.  NC-c.CC=N and NC-c.C=NC are two different cyclic isomers with slightly different geometries and energies.}
  \label{c3n2_table}
  \includegraphics[width=0.9\hsize]{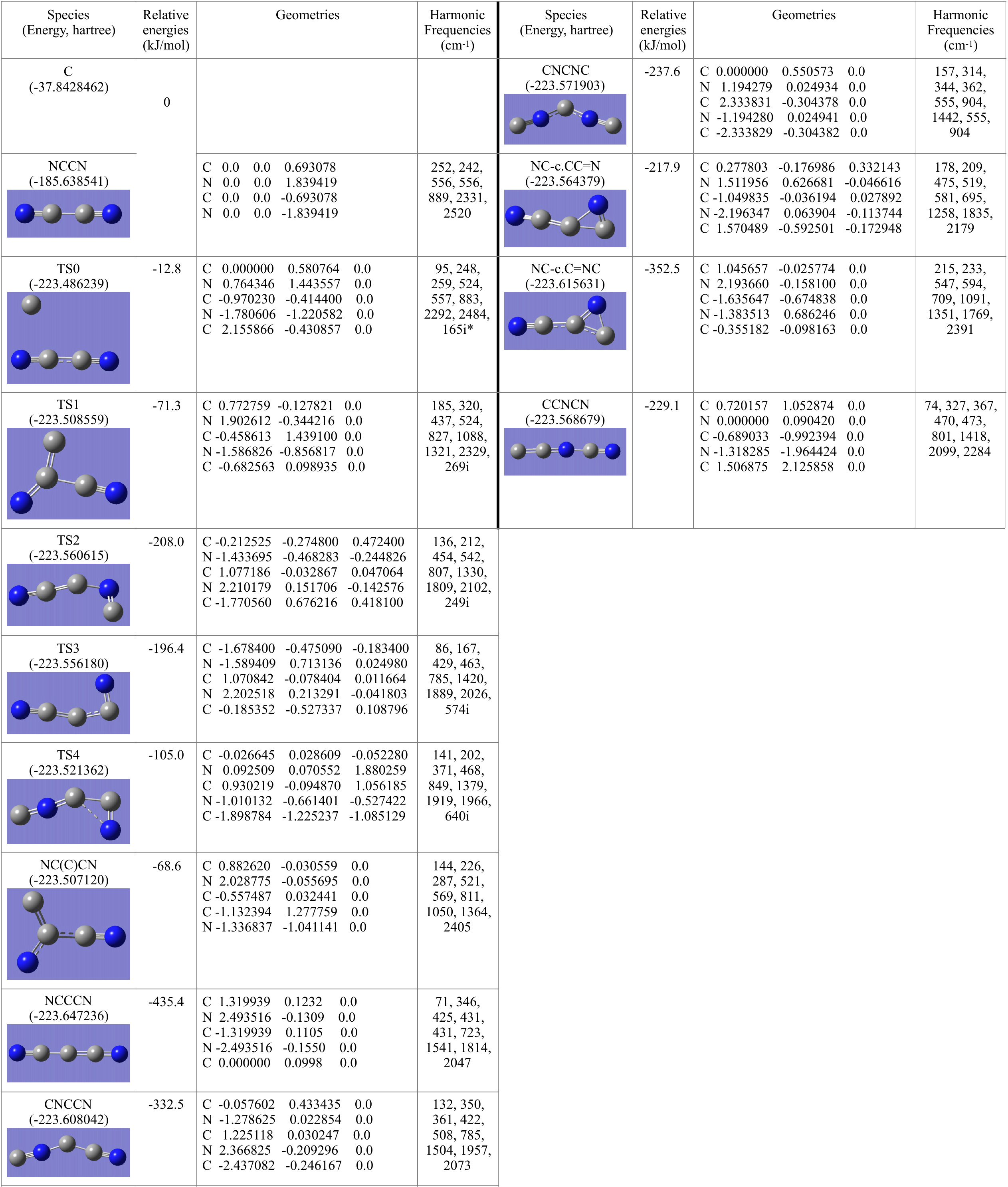}
\end{table*} 

 \begin{figure*}[h]
   \centering
   \includegraphics[width=0.75\hsize]{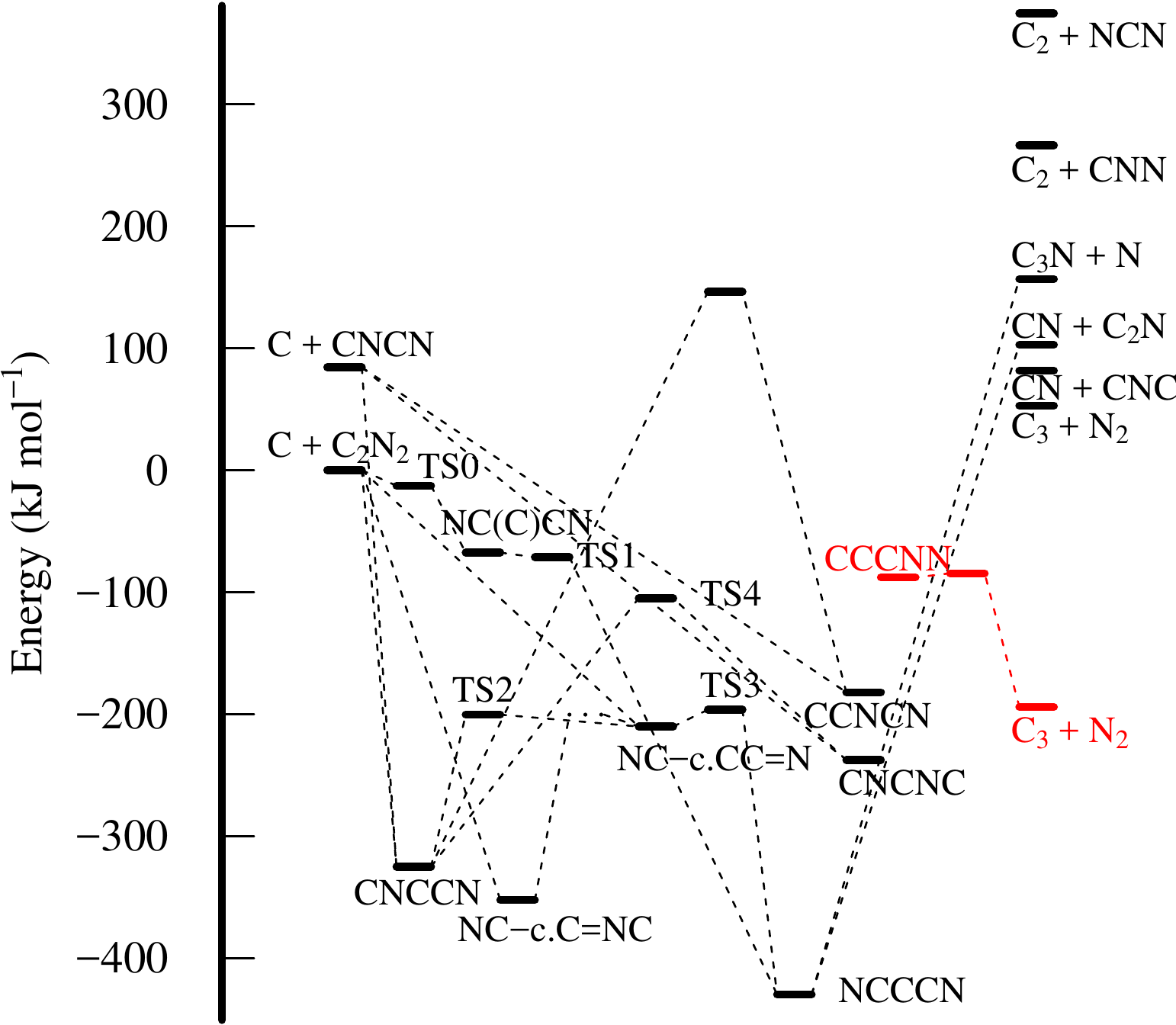}
      \caption{Partial potential energy diagram for the C$_3$N$_2$ system on the triplet surface calculated at the M06-2X/AVTZ level including ZPE. The C$_3$ + N$_2$ exit channel in red is spin forbidden if C$_3$ and N$_2$ are in their singlet ground state. We cannot find the TS from NC-c.CC=N toward NC-c.C=NC at M06-2X/AVTZ level, both species having very similar structure (two different cyclic isomers with slightly different geometries and energies). NC-c.CC=N is a local minimum but directly connected to NC-c.C=NC without barrier at semi-empirical AM1 level. So, this TS is very likely close in energy to NC-c.CC=N.}
   \label{C3N2}
 \end{figure*}

\section{The NCCCN $\rightarrow$ TS1 $\rightarrow$ NC(C)CN isomerization}

 \begin{figure*}[h]
   \centering
   \includegraphics[width=0.9\hsize]{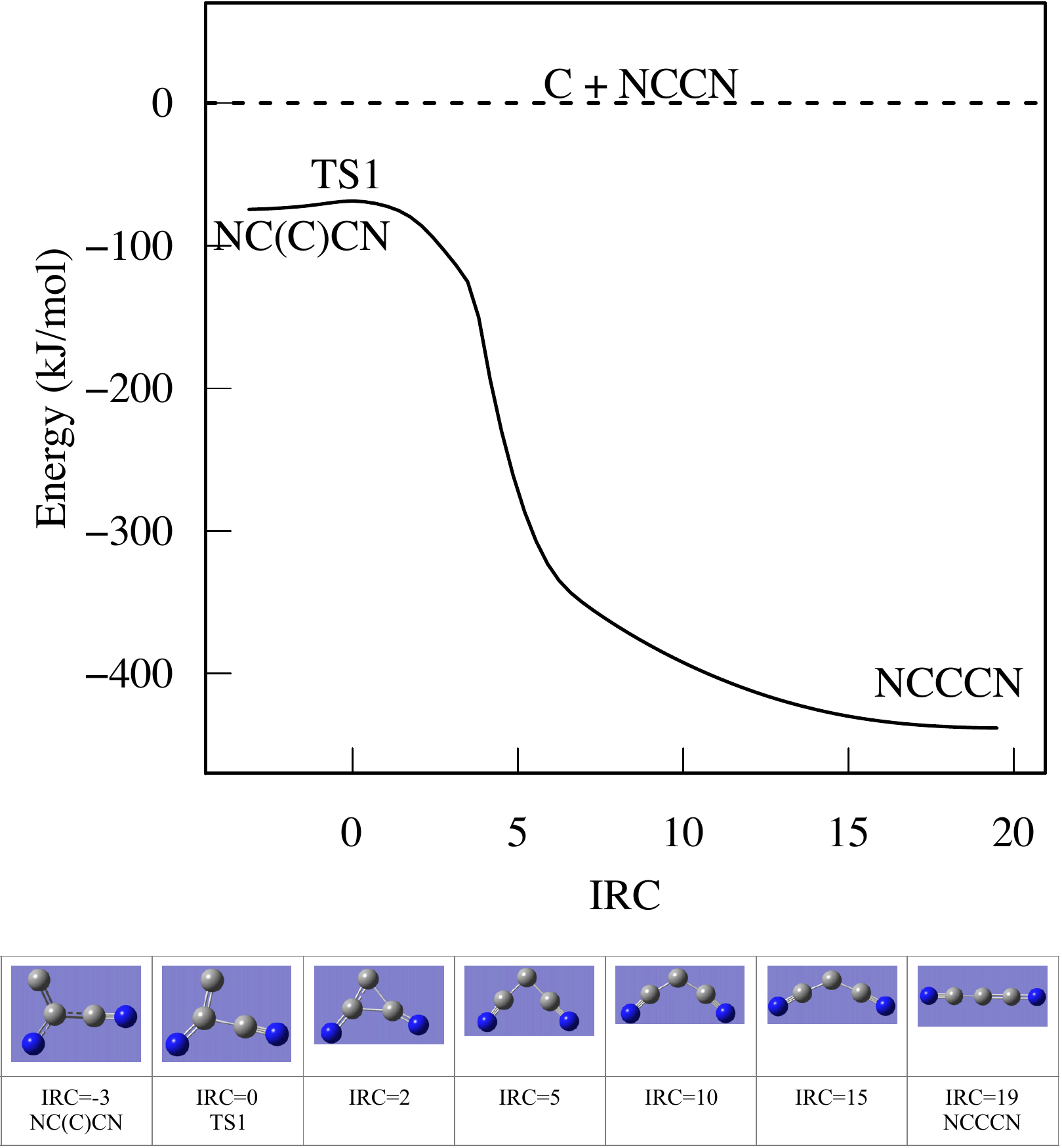}
   \caption{Intrinsic Reaction Coordinate (minimum energy path) at M06-2X/cc-pVTZ for the NCCCN $\rightarrow$ TS1 $\rightarrow$  NC(C)CN isomerization.}
   \label{isomerization1}
 \end{figure*}

\section{The CNCCN  $\rightarrow$  NCCCN isomerization.}

 \begin{figure*}[h]
   \centering
   \includegraphics[width=0.9\hsize]{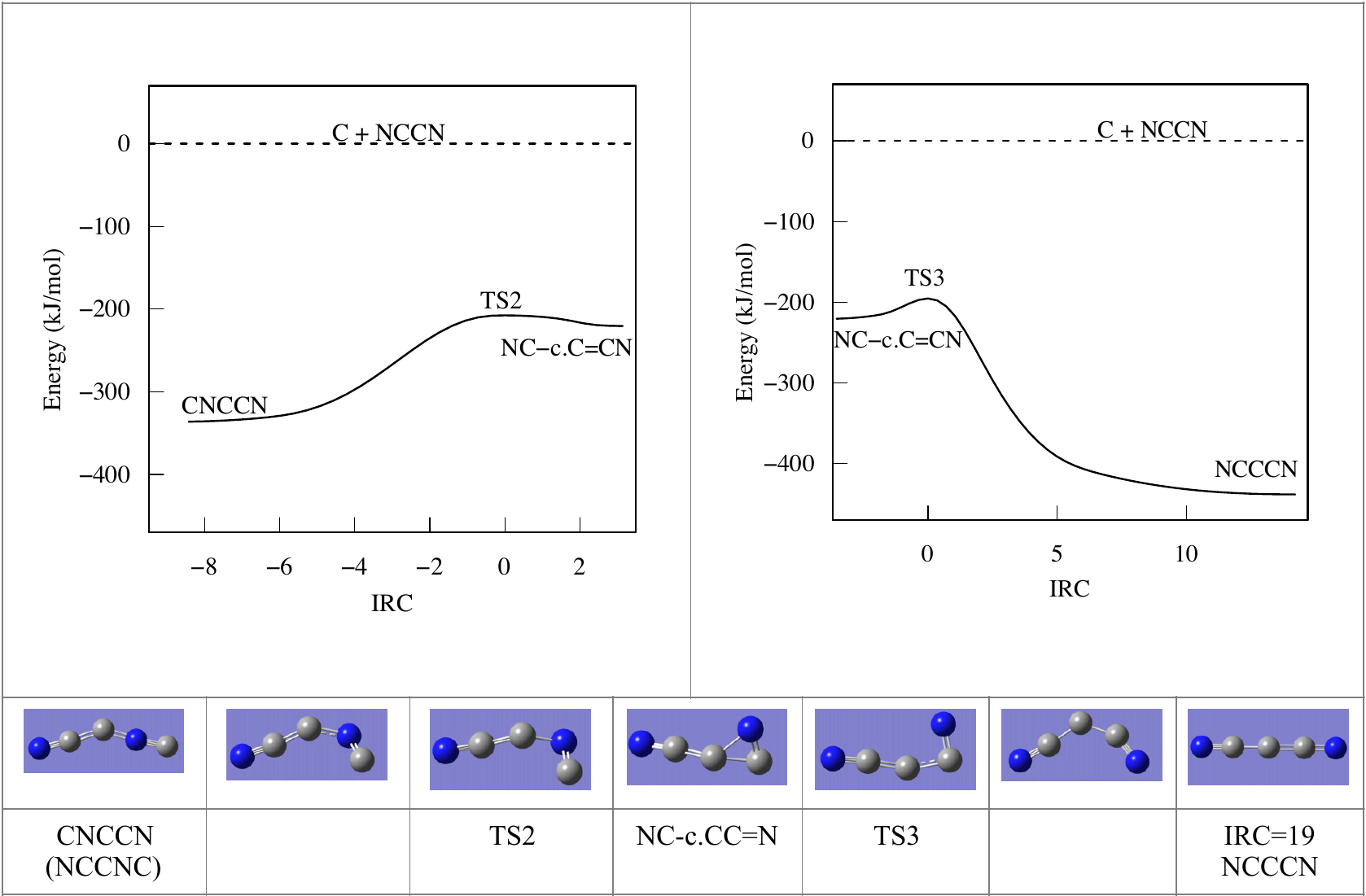}
   \caption{Intrinsic Reaction Coordinate (minimum energy path) at M06-2X/cc-pVTZ for the CNCCN  $\rightarrow$  NCCCN isomerization.}
   \label{isomerization2}
 \end{figure*}

\section{The NCCN $\rightarrow$ CNCN isomerization}

 \begin{figure*}[h]
   \centering
   \includegraphics[width=0.9\hsize]{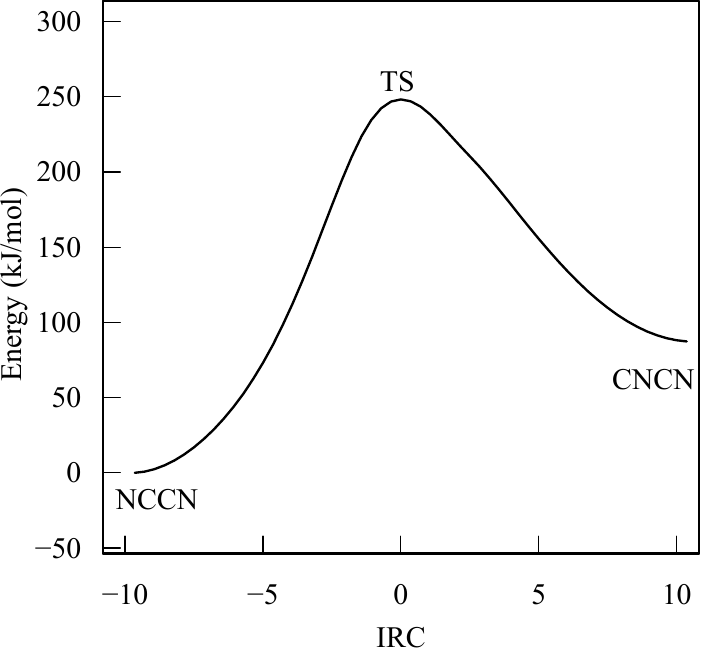}
   \caption{Intrinsic Reaction Coordinate pathway for the NCCN $\rightarrow$ CNCN isomerization calculated at M06-2X/AVTZ level.}
   \label{isomerization3}
 \end{figure*}

\begin{table*}[h]
  \caption{Relative energies at the M06-2X/AVTZ level (in kJ/mol at 0 K including ZPE) with respect to the NCCN energy, geometries and frequencies (in cm$^{-1}$, unscaled, calculated at the M06-2X/AVTZ level) of the various stationary points for the NCCN $\rightarrow$ CNCN isomerization. The absolute energies at the M06-2X/AVTZ level including ZPE in hartree are also given in Column 1. The i label represents an imaginary frequency which is characteristic of transition state. TS is the structure with the highest potential energy along the reaction coordinate.}
  \label{c2n2_cncn}
  \includegraphics[width=0.9\linewidth]{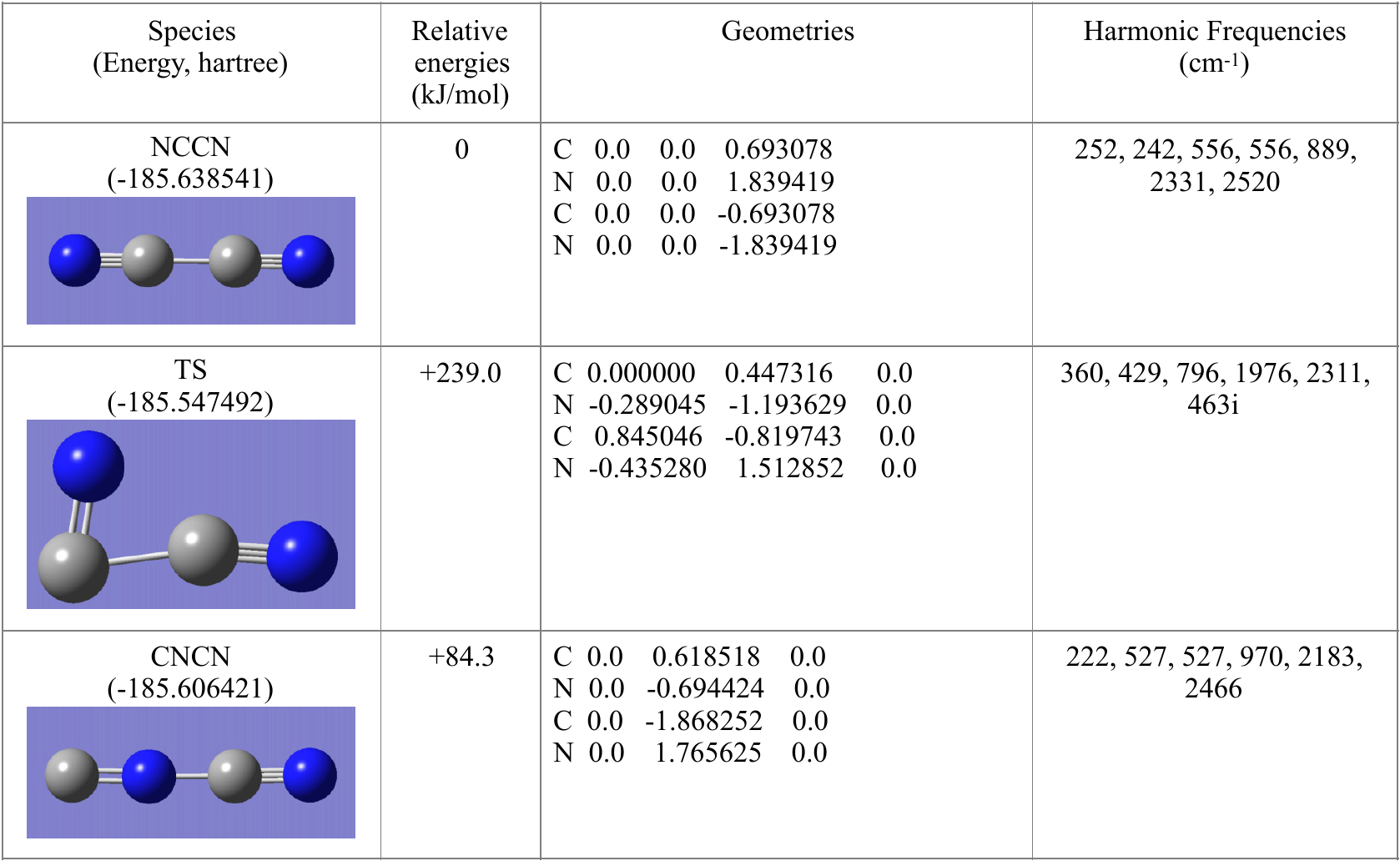}
\end{table*}

\section{N, O + NCCN}

\begin{table*}[h]
  \caption{Relative energies at the M06-2X/AVTZ level (in kJ/mol at 0 K including ZPE) with respect to the N + NCCN and O + NCCN energies, geometries and frequencies (in cm$^{-1}$, unscaled, calculated at the M06-2X/AVTZ level) of the various transition states for N and O atoms additions on NCCN. The i label represents an imaginary frequency which is characteristic of transition state.}
  \label{c2n2}
  \includegraphics[width=0.9\linewidth]{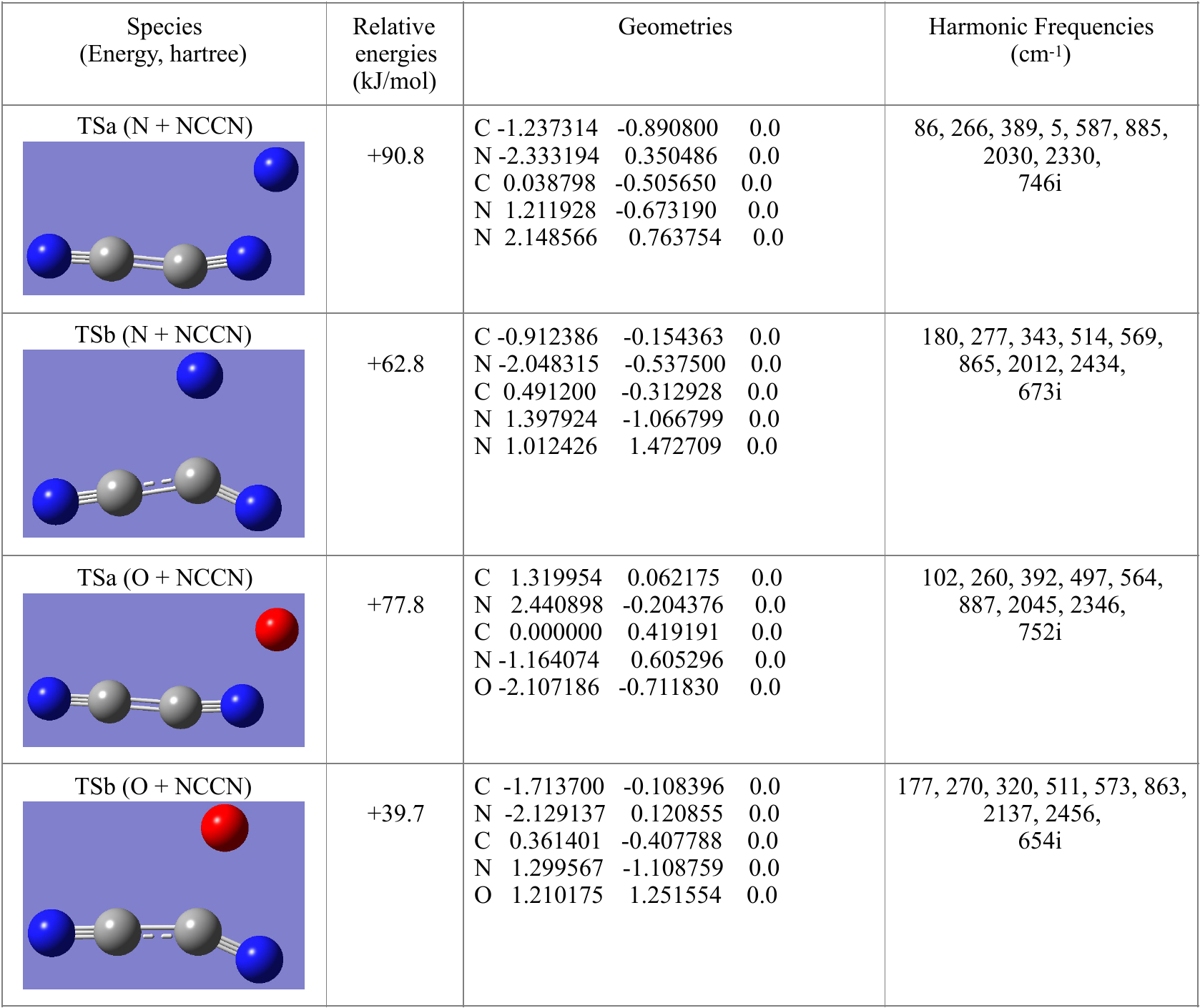}
\end{table*} 

\section{H, N, O + CNCN}

\begin{table*}[h]
  \caption{Relative energies at the M06-2X/AVTZ level (in kJ/mol at 0 K including ZPE) with respect to the H + CNCN, N + CNCN and O + CNCN energies, geometries and frequencies (in cm$^{-1}$, unscaled, calculated at the M06-2X/AVTZ level) of the various transition states for H, N, and O atoms additions on CNCN.  The i label represents an imaginary frequency which is characteristic of transition state.}
  \label{cncn}
  \includegraphics[width=0.9\linewidth]{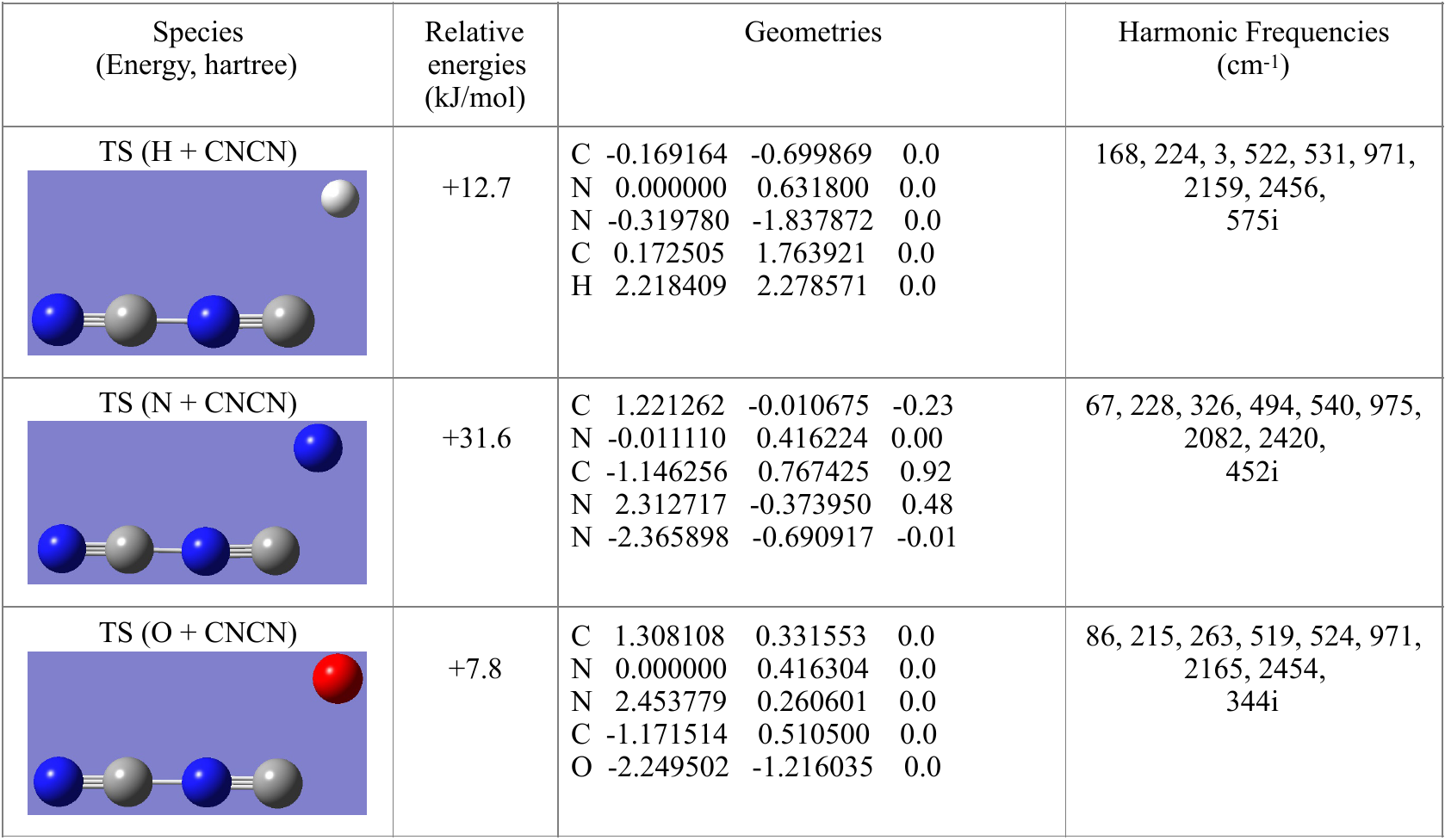}
\end{table*}

\section{The CN + HCNH$^+$ reaction}

\begin{table*}[h]
  \caption{Relative energies at the M06-2X/AVTZ level (in kJ/mol at 0 K including ZPE) with respect to the CN + HCNH$^+$  energy, geometries and frequencies (in cm$^{-1}$, unscaled, calculated at the M06-2X/AVTZ level) of the various stationary points for the CN + HCNH$^+$ reaction. The absolute energies at the M06-2X/AVTZ level including ZPE in hartree are also given in column 1. The i label represents an imaginary frequency which is characteristic of transition state.}
  \label{cn_hcnp+}
  \includegraphics[width=0.9\linewidth]{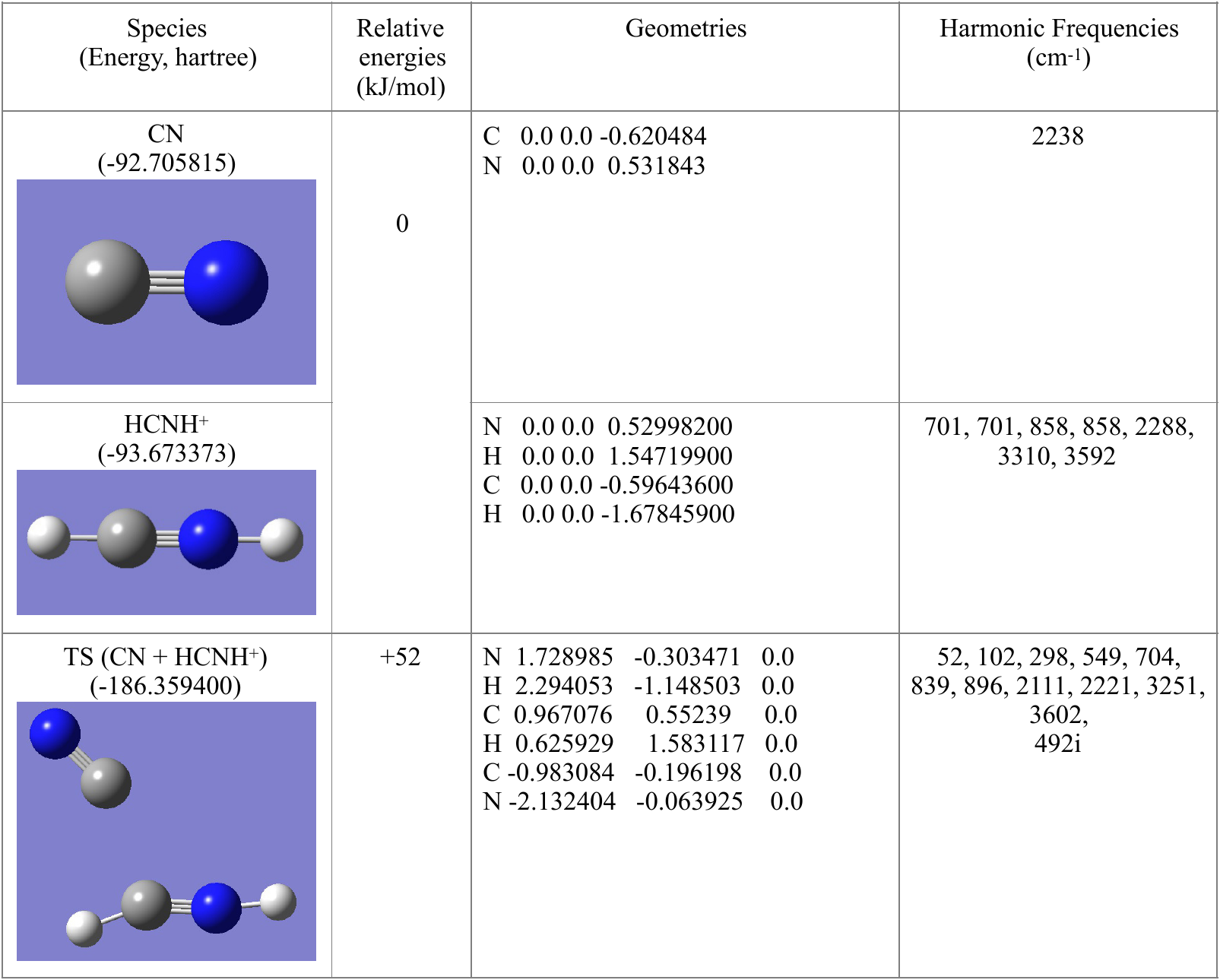}
\end{table*} 

\end{appendix}

%
\bibliographystyle{aa} 
\bibliography{cvastel2019.bbl} 

\begin{thebibliography}{86}
\expandafter\ifx\csname natexlab\endcsname\relax\def\natexlab#1{#1}\fi

\bibitem[{Ag{\'u}ndez {et~al.}(2015)Ag{\'u}ndez, Cernicharo, de~Vicente,
  Marcelino, Roueff, Fuente, Gerin, Gu{\'e}lin, Albo, Barcia, Barbas,
  Bola{\~n}o, Colomer, Diez, Gallego, G{\'o}mez-Gonz{\'a}lez,
  L{\'o}pez-Fern{\'a}ndez, L{\'o}pez-Fern{\'a}ndez, L{\'o}pez-P{\'e}rez, Malo,
  Serna, \& Tercero}]{agundez2015}
Ag{\'u}ndez, M., Cernicharo, J., de~Vicente, P., {et~al.} 2015, Astronomy and
  Astrophysics, 579, L10

\bibitem[{Ag{\'u}ndez {et~al.}(2018)Ag{\'u}ndez, Marcelino, \&
  Cernicharo}]{agundez2018b}
Ag{\'u}ndez, M., Marcelino, N., \& Cernicharo, J. 2018, The Astrophysical
  Journal Letters, 861

\bibitem[{{Ag{\'u}ndez} {et~al.}(2018){Ag{\'u}ndez}, {Marcelino}, {Cernicharo},
  \& {Tafalla}}]{agundez2018}
{Ag{\'u}ndez}, M., {Marcelino}, N., {Cernicharo}, J., \& {Tafalla}, M. 2018,
  Astronomy and Astrophysics, 611, L1

\bibitem[{Amano \& Scappini(1991)}]{amano1991}
Amano, T. \& Scappini, F. 1991, Journal of Chemical Physics, 95, 2280

\bibitem[{Anicich(2003)}]{anicich2003}
Anicich, V.~G. 2003, JPL publication, 03-19 NASA

\bibitem[{Bockelee-Morvan \& Crovisier(1985)}]{bockelee1985}
Bockelee-Morvan, D. \& Crovisier, J. 1985, Astronomy and Astrophysics, 151, 90

\bibitem[{Bockel{\'e}e-Morvan {et~al.}(2000)Bockel{\'e}e-Morvan, Lis, Wink,
  Despois, Crovisier, \& Bachiller}]{bockelee2000}
Bockel{\'e}e-Morvan, D., Lis, D.~C., Wink, J.~E., {et~al.} 2000, Astronomy and
  Astrophysics, 353, 1101

\bibitem[{Bonev \& Komitov(2000)}]{bonev2000}
Bonev, B. \& Komitov, B. 2000, Bulletin of the American Astronomical Society,
  32, 1072

\bibitem[{Caselli {et~al.}(2012)Caselli, Keto, Bergin, Tafalla, Aikawa,
  Douglas, Pagani, Y{\'\i}ld{\'\i}z, van~der Tak, Walmsley, Codella, Nisini,
  Kristensen, \& van Dishoeck}]{caselli2012}
Caselli, P., Keto, E., Bergin, E.~A., {et~al.} 2012, The Astrophysical Journal
  Letters, 759, L37

\bibitem[{Cazzoli \& Puzzarini(2006)}]{cazzoli2006}
Cazzoli, G. \& Puzzarini, C. 2006, Journal of Molecular Spectroscopy, 240, 153

\bibitem[{Cernicharo \& Guelin(1987)}]{cernicharo1987}
Cernicharo, J. \& Guelin, M. 1987, Astronomy and Astrophysics, 176, 299

\bibitem[{Crutcher {et~al.}(1984)Crutcher, Churchwell, \&
  Ziurys}]{crutcher1984}
Crutcher, R.~M., Churchwell, E., \& Ziurys, L.~M. 1984, The Astrophysical
  Journal, 283, 668

\bibitem[{Dutuit {et~al.}(2013)Dutuit, Carrasco, Thissen, Vuitton, Christian,
  Pascal, \& Nadia}]{dutuit2013}
Dutuit, O., Carrasco, N., Thissen, R., {et~al.} 2013, The Astrophysical Journal
  Supplement Series, 204, 20

\bibitem[{Ehrenfreund \& Charnley(2000)}]{ehrenfreund2000}
Ehrenfreund, P. \& Charnley, S.~B. 2000, Annual Review of Astronomy and
  Astrophysics, 38, 427

\bibitem[{Feh{\'e}r {et~al.}(2016)Feh{\'e}r, T{\'o}th, Ward-Thompson, Kirk,
  Kraus, Pelkonen, Pint{\'e}r, \& Zahorecz}]{feher2016}
Feh{\'e}r, O., T{\'o}th, L.~V., Ward-Thompson, D., {et~al.} 2016, Astronomy and
  Astrophysics, 590, 75

\bibitem[{Florescu-Mitchell \& Mitchell(2006)}]{florescumitchell2006}
Florescu-Mitchell, A.~I. \& Mitchell, J. B.~A. 2006, Phys. Rep., 430, 277

\bibitem[{Fray {et~al.}(2005)Fray, B{\'e}nilan, Cottin, Gazeau, \&
  Crovisier}]{fray2005}
Fray, N., B{\'e}nilan, Y., Cottin, H., Gazeau, M.-C., \& Crovisier, J. 2005,
  Planetary and Space Science, 53, 1243

\bibitem[{Geiss {et~al.}(1999)Geiss, Altwegg, Balsiger, \& Graf}]{geiss1999}
Geiss, J., Altwegg, K., Balsiger, H., \& Graf, S. 1999, Space Science Reviews,
  90, 253

\bibitem[{Geppert {et~al.}(2004)Geppert, Ehlerding, Hellberg, Semaniak,
  {\"O}sterdahl, Kami{\'n}ska, Al-Khalili, Zhaunerchyk, Thomas, af~Ugglas,
  K{\"a}llberg, Simonsson, \& Larsson}]{geppert2004}
Geppert, W.~D., Ehlerding, A., Hellberg, F., {et~al.} 2004, The Astrophysical
  Journal, 613, 1302

\bibitem[{Gerry {et~al.}(1990)Gerry, Stroh, \& Winnewisser}]{gerry1990}
Gerry, M. C.~L., Stroh, F., \& Winnewisser, M. 1990, Journal of Molecular
  Spectroscopy, 140, 147

\bibitem[{Gottlieb {et~al.}(1983)Gottlieb, Gottlieb, Thaddeus, \&
  Kawamura}]{gottlieb1983}
Gottlieb, C.~A., Gottlieb, E.~W., Thaddeus, P., \& Kawamura, H. 1983, The
  Astrophysical Journal, 275, 916

\bibitem[{Gu{\'e}lin {et~al.}(1998)Gu{\'e}lin, Neininger, \&
  Cernicharo}]{guelin1998}
Gu{\'e}lin, M., Neininger, N., \& Cernicharo, J. 1998, Astronomy and
  Astrophysics, 335, L1

\bibitem[{Hebrard {et~al.}(2013)Hebrard, Dobrijevic, Loison, Bergeat, Hickson,
  \& Caralp}]{hebrard2013}
Hebrard, E., Dobrijevic, M., Loison, J.~C., {et~al.} 2013, Astronomy and
  Astrophysics, 552

\bibitem[{Herbst {et~al.}(2000)Herbst, Terzieva, \& Talbi}]{herbst2000}
Herbst, E., Terzieva, R., \& Talbi, D. 2000, MNRAS, 311, 869

\bibitem[{{Hickson} {et~al.}(2016){Hickson}, {Loison}, \&
  {Wakelam}}]{hickson2016}
{Hickson}, K.~M., {Loison}, J.-C., \& {Wakelam}, V. 2016, Chemical Physics
  Letters, 659, 70

\bibitem[{Hily-Blant {et~al.}(2018)Hily-Blant, Faure, Vastel, Magalhaes,
  Lefloch, \& Bachiller}]{hily-blant2018}
Hily-Blant, P., Faure, A., Vastel, C., {et~al.} 2018, Monthly Notices of the
  Royal Astronomical Society, 480, 1174

\bibitem[{Hily-Blant {et~al.}(2008)Hily-Blant, Walmsley, Pineau Des~For{\^e}ts,
  \& Flower}]{hily-blant2008}
Hily-Blant, P., Walmsley, M., Pineau Des~For{\^e}ts, G., \& Flower, D. 2008,
  Astronomy and Astrophysics, 480, 5

\bibitem[{Hirota {et~al.}(1998)Hirota, Yamamoto, Mikami, \&
  Ohishi}]{hirota1998}
Hirota, T., Yamamoto, S., Mikami, H., \& Ohishi, M. 1998, The Astrophysical
  Journal, 503, 717

\bibitem[{Irvine {et~al.}(1988)Irvine, Friberg, Hjalmarson, Ishikawa, Kaifu, \&
  Kawaguchi}]{irvine1988}
Irvine, W.~M., Friberg, P., Hjalmarson, A., {et~al.} 1988, The Astrophysical
  Journal, 334, 107

\bibitem[{{Jenkins}(2009)}]{jenkins2009}
{Jenkins}, E.~B. 2009, The Astrophysical Journal, 700, 1299

\bibitem[{Jim{\'e}nez-Serra {et~al.}(2016)Jim{\'e}nez-Serra, Vasyunin, Caselli,
  Marcelino, Billot, Viti, Testi, Vastel, Lefloch, \&
  Bachiller}]{jimenez-serra2016}
Jim{\'e}nez-Serra, I., Vasyunin, A.~I., Caselli, P., {et~al.} 2016, The
  Astrophysical Journal Letters, 830, L6

\bibitem[{Kawaguchi {et~al.}(1994)Kawaguchi, Kasai, Ishikawa, Ohishi, Kaifu, \&
  Amano}]{kawaguchi1994}
Kawaguchi, K., Kasai, Y., Ishikawa, S.-I., {et~al.} 1994, The Astrophysical
  Journal Letters, 420, L95

\bibitem[{Kawaguchi {et~al.}(1992{\natexlab{a}})Kawaguchi, Ohishi, Ishikawa, \&
  Kaifu}]{kawaguchi1992a}
Kawaguchi, K., Ohishi, M., Ishikawa, S.-I., \& Kaifu, N. 1992{\natexlab{a}},
  The Astrophysical Journal, 386, L51

\bibitem[{Kawaguchi {et~al.}(1992{\natexlab{b}})Kawaguchi, Takano, Ohishi,
  Ishikawa, Miyazawa, Kaifu, Yamashita, Yamamoto, Saito, Ohshima, \&
  Endo}]{kawaguchi1992b}
Kawaguchi, K., Takano, S., Ohishi, M., {et~al.} 1992{\natexlab{b}}, \apj, 396,
  L49

\bibitem[{Keto {et~al.}(2014)Keto, Rawlings, \& Caselli}]{keto2014}
Keto, E., Rawlings, J., \& Caselli, P. 2014, Monthly Notices of the Royal
  Astronomical Society, 440, 2616

\bibitem[{Kolos \& Grabowski(2000)}]{kolos2000}
Kolos, R. \& Grabowski, Z. 2000, Astrophysics and Space Science, 271, 65

\bibitem[{Kuiper {et~al.}(1996)Kuiper, Langer, \& Velusamy}]{kuiper1996}
Kuiper, T. B.~H., Langer, W.~D., \& Velusamy, T. 1996, The Astrophysical
  Journal, 468, 761

\bibitem[{{Lefloch} {et~al.}(2018){Lefloch}, {Bachiller}, {Ceccarelli},
  {Cernicharo}, {Codella}, {Fuente}, {Kahane}, {L{\'o}pez-Sepulcre}, {Tafalla},
  {Vastel}, {Caux}, {Gonz{\'a}lez-Garc{\'{\i}}a}, {Bianchi}, {G{\'o}mez-Ruiz},
  {Holdship}, {Mendoza}, {Ospina-Zamudio}, {Podio}, {Qu{\'e}nard}, {Roueff},
  {Sakai}, {Viti}, {Yamamoto}, {Yoshida}, {Favre}, {Monfredini},
  {Quiti{\'a}n-Lara}, {Marcelino}, {Boechat-Roberty}, \&
  {Cabrit}}]{lefloch2018}
{Lefloch}, B., {Bachiller}, R., {Ceccarelli}, C., {et~al.} 2018, accepted for
  publication in MNRAS

\bibitem[{{Loison} {et~al.}(2016){Loison}, {Ag{\'u}ndez}, {Marcelino},
  {Wakelam}, {Hickson}, {Cernicharo}, {Gerin}, {Roueff}, \&
  {Gu{\'e}lin}}]{loison2016}
{Loison}, J.-C., {Ag{\'u}ndez}, M., {Marcelino}, N., {et~al.} 2016, Monthly
  Notices of the Royal Astronomical Society, 456, 4101

\bibitem[{Loison {et~al.}(2017)Loison, Ag{\'u}ndez, Wakelam, Roueff, Gratier,
  Marcelino, Reyes, Cernicharo, \& Gerin}]{loison2017}
Loison, J.-C., Ag{\'u}ndez, M., Wakelam, V., {et~al.} 2017, Monthly Notices of
  the Royal Astronomical Society, 470, 4075

\bibitem[{Loison \& Hickson(2015)}]{loison2015}
Loison, J.~C. \& Hickson, K.~M. 2015, Chemical Physics Letters, 635, 174

\bibitem[{Loison {et~al.}(2014)Loison, Wakelam, Hickson, Bergeat, \&
  Mereau}]{loison2014}
Loison, J.-C., Wakelam, V., Hickson, K.~M., Bergeat, A., \& Mereau, R. 2014,
  Monthly Notices of the Royal Astronomical Society, 437, 930

\bibitem[{McGonagle \& Irvine(1996)}]{mcgonagle1996}
McGonagle, D. \& Irvine, W. 1996, Astronomy and Astrophysics, 310, 970

\bibitem[{Michael(1980)}]{michael1980}
Michael, J.~V. 1980, Chemical Physics Letters, 96, 129

\bibitem[{Minh {et~al.}(1993)Minh, Irvine, Ohishi, Ishikawa, Saito, \&
  Kaifu}]{minh1993}
Minh, Y., Irvine, W., Ohishi, M., {et~al.} 1993, Astronomy and Astrophysics,
  267, 229

\bibitem[{M{\"u}ller {et~al.}(2008)M{\"u}ller, Belloche, Menten, Comito, \&
  Schilke}]{muller2008}
M{\"u}ller, H. S.~P., Belloche, A., Menten, K.~M., Comito, C., \& Schilke, P.
  2008, Journal of Molecular Spectroscopy, 251, 319

\bibitem[{M{\"u}ller {et~al.}(2005)M{\"u}ller, Schl{\"o}der, Stutzki, \&
  Winnewisser}]{muller2005}
M{\"u}ller, H. S.~P., Schl{\"o}der, F., Stutzki, J., \& Winnewisser, G. 2005,
  Journal of Molecular Structure, 742, 215

\bibitem[{Ohashi {et~al.}(1999)Ohashi, Lee, Wilner, \& Hayashi}]{ohashi1999}
Ohashi, N., Lee, S., Wilner, D., \& Hayashi, M. 1999, The Astrophysical
  Journal, 518, 41

\bibitem[{Ohishi {et~al.}(1994)Ohishi, McGonagle, Irvine, \&
  Saito}]{ohishi1994}
Ohishi, M., McGonagle, D., Irvine, William M.;~andYamamoto, S., \& Saito, S.
  1994, The Astrophysical Journal Letters, 427, 51

\bibitem[{Osamura {et~al.}(1999)Osamura, Fukuzawa, Terzieva, \&
  Herbst}]{osamura1999}
Osamura, Y., Fukuzawa, K., Terzieva, R., \& Herbst, E. 1999, The Astrophysical
  Journal, 519, 697

\bibitem[{Petrie {et~al.}(2003)Petrie, Millar, \& Markwick}]{petrie2003}
Petrie, S., Millar, T.~J., \& Markwick, A.~J. 2003, Monthly Notices of the
  Royal Astronomical Society, 341, 609

\bibitem[{Petrie \& Osamura(2004)}]{petrie2004}
Petrie, S. \& Osamura, Y. 2004, Journal of Physical Chemistry A, 108, 3623

\bibitem[{Plessis {et~al.}(2012)Plessis, Carrasco, Dobrijevic, \&
  Pernot}]{plessis2012}
Plessis, S., Carrasco, N., Dobrijevic, M., \& Pernot, P. 2012, Icarus, 219, 254

\bibitem[{{Qu{\'e}nard} {et~al.}(2017{\natexlab{a}}){Qu{\'e}nard},
  {Bottinelli}, \& {Caux}}]{quenard2017b}
{Qu{\'e}nard}, D., {Bottinelli}, S., \& {Caux}, E. 2017{\natexlab{a}}, Monthly
  Notices of the Royal Astronomical Society, 468, 685

\bibitem[{{Qu{\'e}nard} {et~al.}(2017{\natexlab{b}}){Qu{\'e}nard}, {Vastel},
  {Ceccarelli}, {Hily-Blant}, {Lefloch}, \& {Bachiller}}]{quenard2017a}
{Qu{\'e}nard}, D., {Vastel}, C., {Ceccarelli}, C., {et~al.} 2017{\natexlab{b}},
  \mnras, 470, 3194

\bibitem[{Raksit \& Bohme(1984)}]{raksit1984}
Raksit, A.~B. \& Bohme, D.~K. 1984, Canadian Journal of Chemistry, 62, 2123

\bibitem[{Reiter \& Janev(2010)}]{reiter2010}
Reiter, D. \& Janev, R.~K. 2010, Contrib. Plasm. Phys., 50, 986

\bibitem[{Rivilla {et~al.}(2018)Rivilla, Mart{\'\i}n-Pintado,
  Jim{\'e}nez-Serra, Zeng, Mart{\'\i}n, Armijos-Abenda{\~n}o, Requena-Torres,
  Aladro, \& Riquelme}]{rivilla2018}
Rivilla, V.~M., Mart{\'\i}n-Pintado, J., Jim{\'e}nez-Serra, I., {et~al.} 2018,
  Monthly Notices of the Royal Astronomical Society: Letters

\bibitem[{Ruaud {et~al.}(2016)Ruaud, Wakelam, \& Hersant}]{ruaud2016}
Ruaud, M., Wakelam, V., \& Hersant, F. 2016, Monthly Notices of the Royal
  Astronomical Society, 459, 3756

\bibitem[{Safrany \& Jaster(1968{\natexlab{a}})}]{safrany1968c}
Safrany, D. \& Jaster, W. 1968{\natexlab{a}}, The Journal of Physical Chemistry
  A, 72, 3305

\bibitem[{Safrany \& Jaster(1968{\natexlab{b}})}]{safrany1968a}
Safrany, D. \& Jaster, W. 1968{\natexlab{b}}, The Journal of Chemical Physics,
  72, 3305

\bibitem[{Safrany \& Jaster(1968{\natexlab{c}})}]{safrany1968b}
Safrany, D.~R. \& Jaster, W. 1968{\natexlab{c}}, The Journal of Chemical
  Physics, 72, 3318

\bibitem[{Schilke {et~al.}(1991)Schilke, Walmsley, Henkel, \&
  Millar}]{schilke1991}
Schilke, P., Walmsley, C.~M., Henkel, C., \& Millar, T.~J. 1991, Astronomy and
  Astrophysics, 247, 487

\bibitem[{Shivani \& Tandon(2017)}]{shivani2017}
Shivani, M. \& Tandon, P. 2017, Research in Astronomy and Astrophysics, 17, 1

\bibitem[{{Spezzano} {et~al.}(2016){Spezzano}, {Bizzocchi}, {Caselli}, {Harju},
  \& {Br{\"u}nken}}]{spezzano2016}
{Spezzano}, S., {Bizzocchi}, L., {Caselli}, P., {Harju}, J., \& {Br{\"u}nken},
  S. 2016, Astronomy and Astrophysics, 592, 11

\bibitem[{Swings \& Haser(1956)}]{swings1956}
Swings, P. \& Haser, L. 1956, Institut d'Astrophysique, Li{\`e}ge

\bibitem[{Tafalla {et~al.}(1998)Tafalla, Mardones, Myers, Caselli, Bachiller,
  \& Benson}]{tafalla1998}
Tafalla, M., Mardones, D., Myers, P.~C., {et~al.} 1998, The Astrophysical
  Journal, 504, 900

\bibitem[{{Tafalla} {et~al.}(2006){Tafalla}, {Santiago-Garc{\'{\i}}a}, {Myers},
  {Caselli}, {Walmsley}, \& {Crapsi}}]{tafalla2006}
{Tafalla}, M., {Santiago-Garc{\'{\i}}a}, J., {Myers}, P.~C., {et~al.} 2006, The
  Astrophysical Journal, 455, 577

\bibitem[{Takano {et~al.}(1998)Takano, Masuda, Hirahara, Suzuki, Ohishi,
  Ishikawa, Kaifu, Kasai, Kawaguchi, \& Wilson}]{takano1998}
Takano, S., Masuda, A., Hirahara, Y., {et~al.} 1998, Astronomy and
  Astrophysics, 329, 1156

\bibitem[{Teanby {et~al.}(2006)Teanby, Irwin, de~Kok, Nixon, Coustenis, \&
  B{\'e}zard}]{teanby2006}
Teanby, N.~A., Irwin, P. G.~J., de~Kok, R., {et~al.} 2006, Icarus, 181, 243

\bibitem[{Turner {et~al.}(1999)Turner, Terzieva, \& Herbst}]{turner1999}
Turner, B.~E., Terzieva, R., \& Herbst, E. 1999, The Astrophysical Journal,
  518, 699

\bibitem[{Vastel {et~al.}(2015{\natexlab{a}})Vastel, Bottinelli, Caux, Glorian,
  \& Boiziot}]{vastel2015a}
Vastel, C., Bottinelli, S., Caux, E., Glorian, J.-M., \& Boiziot, M.
  2015{\natexlab{a}}, in SF2A-2015: Proceedings of the Annual meeting of the
  French Society of Astronomy and Astrophysics. Eds.: F. Martins, S. Boissier,
  V. Buat, L. Cambr{\'e}sy, P. Petit, pp.313-316, 313--316

\bibitem[{Vastel {et~al.}(2014)Vastel, Ceccarelli, Lefloch, \&
  Bachiller}]{vastel2014}
Vastel, C., Ceccarelli, C., Lefloch, B., \& Bachiller, R. 2014, The
  Astrophysical Journal Letters, 795, L2

\bibitem[{{Vastel} {et~al.}(2018a){Vastel}, {Kawaguchi}, {Qu{\'e}nard},
  {Ohishi}, {Lefloch}, {Bachiller}, \& {M{\"u}ller}}]{vastel2018a}
{Vastel}, C., {Kawaguchi}, K., {Qu{\'e}nard}, D., {et~al.} 2018a, Monthly
  Notices of the Royal Astronomical Society: Letters

\bibitem[{Vastel {et~al.}(2018b)Vastel, Qu{\'e}nard, Le~Gal, Wakelam,
  Andrianasolo, Caselli, Vidal, Ceccarelli, Lefloch, \&
  Bachiller}]{vastel2018b}
Vastel, C., Qu{\'e}nard, D., Le~Gal, R., {et~al.} 2018b, Monthly Notices of the
  Royal Astronomical Society, 478, 5514

\bibitem[{Vastel {et~al.}(2015{\natexlab{b}})Vastel, Yamamoto, Lefloch, \&
  Bachiller}]{vastel2015b}
Vastel, C., Yamamoto, S., Lefloch, B., \& Bachiller, R. 2015{\natexlab{b}},
  Astronomy and Astrophysics, 582, L3

\bibitem[{{Vidal} {et~al.}(2017){Vidal}, {Loison}, {Jaziri}, {Ruaud},
  {Gratier}, \& {Wakelam}}]{vidal2017}
{Vidal}, T.~H.~G., {Loison}, J.-C., {Jaziri}, A.~Y., {et~al.} 2017, Monthly
  Notices of the Royal Astronomical Society, 469, 435

\bibitem[{{Vigren} {et~al.}(2012){Vigren}, {Semaniak}, {Hamberg},
  {Zhaunerchyk}, {Kaminska}, {Thomas}, {af Ugglas}, {Larsson}, \&
  {Geppert}}]{vigren2012}
{Vigren}, E., {Semaniak}, J., {Hamberg}, M., {et~al.} 2012, Planetary and Space
  Science, 60, 102

\bibitem[{Wakelam {et~al.}(2012)Wakelam, Herbst, Loison, Smith, Chandrasekaran,
  Pavone, Adams, Bacchus-Montabonel, Bergeat, B{\'e}roff, Bierbaum, Chabot,
  Dalgarno, van Dishoeck, Faure, Geppert, Gerlich, Galli, H{\'e}brard, Hersant,
  Hickson, Honvault, Klippenstein, Le~Picard, Nyman, Pernot, Schlemmer, Selsis,
  Sims, Talbi, Tennyson, Troe, Wester, \& Wiesenfeld}]{wakelam2012}
Wakelam, V., Herbst, E., Loison, J.-C., {et~al.} 2012, The Astrophysical
  Journal Supplement Series, 199, 21

\bibitem[{{Wakelam} {et~al.}(2006){Wakelam}, {Herbst}, \&
  {Selsis}}]{wakelam2006}
{Wakelam}, V., {Herbst}, E., \& {Selsis}, F. 2006, Astronomy and Astrophysics,
  451, 551

\bibitem[{Wakelam {et~al.}(2015)Wakelam, Loison, Herbst, Pavone, Bergeat,
  B{\'e}roff, Chabot, Faure, Galli, Geppert, Gerlich, Gratier, Harada, Hickson,
  Honvault, Klippenstein, Le~Picard, Nyman, Ruaud, Schlemmer, Sims, Talbi,
  Tennyson, \& Wester}]{wakelam2015a}
Wakelam, V., Loison, J.-C., Herbst, E., {et~al.} 2015, The Astrophysical
  Journal Supplement Series, 217, 20

\bibitem[{{Wakelam} {et~al.}(2017){Wakelam}, {Loison}, {Mereau}, \&
  {Ruaud}}]{wakelam2017}
{Wakelam}, V., {Loison}, J.-C., {Mereau}, R., \& {Ruaud}, M. 2017, Molecular
  Astrophysics, 6, 22

\bibitem[{Wakelam {et~al.}(2010)Wakelam, Smith, Herbst, Troe, \&
  Geppert}]{wakelam2010}
Wakelam, V., Smith, I. W.~M., Herbst, E., Troe, J., \& Geppert, W. e.~a. 2010,
  Space Science Review, 156, 13

\bibitem[{Whyte \& Phillips(1983)}]{whyte1983}
Whyte, A.~R. \& Phillips, L.~F. 1983, Chemical Physics Letters, 98, 590

\bibitem[{Yamamoto \& Saito(1992)}]{yamamoto1992}
Yamamoto, S. \& Saito, S. 1992, Journal of Chemical Physics, 96, 4157

\bibitem[{Zhao \& Truhlar(2008)}]{zhao2008}
Zhao, Y. \& Truhlar, D. 2008, Theor. Chem. Acc., 120, 215

\end{thebibliography}

%

\end{document}